\documentclass[fleqn,usenatbib]{mnras}
\usepackage{hyperref}
\usepackage{newtxtext}
\usepackage[T1]{fontenc}
\usepackage{graphicx}	
\usepackage{amsmath}	

\usepackage{ae,aecompl}
\usepackage{comment}

\usepackage{amssymb}	
\usepackage{verbatim}
\usepackage{natbib}
\usepackage{etoolbox}
\usepackage{longtable}
\usepackage{latexsym}

\usepackage{ragged2e}

\DeclareRobustCommand{\VAN}[3]{#2}
\let\VANthebibliography\thebibliography
\def\thebibliography{\DeclareRobustCommand{\VAN}[3]{##3}\VANthebibliography}

\usepackage{chemformula}

\DeclareSymbolFont{UPM}{U}{eur}{m}{n}
\DeclareMathSymbol{\newmu}{0}{UPM}{"16}
\newcommand\micro{\ifmmode\newmu\else$\newmu$\fi}
\renewcommand\micron{\micro m}
\newcommand\microns{\micron}

\let\oldsim=\sim
\renewcommand\sim{\ifmmode\oldsim\else\math{\oldsim}\fi}
\let\oldpm=\pm
\renewcommand\pm{\ifmmode\oldpm\else\math{\oldpm}\fi}
\newcommand\by{\ifmmode\times\else\math{\times}\fi}

\newbox{\wdbox}

\newcommand\added[1]{{\if\trackchanges1\bf\fi{#1}}}

\catcode`@=11

\renewcommand\math[1]{$#1$}

\let\oldmsp=\sp
\let\oldmsb=\sb
\def\sp#1{\ifmmode
           \oldmsp{#1}%
         \else\strut\raise.85ex\hbox{\scriptsize #1}\fi}
\def\sb#1{\ifmmode
           \oldmsb{#1}%
         \else\strut\raise-.54ex\hbox{\scriptsize #1}\fi}
\newbox\@sp
\newbox\@sb

\newcommand\JWST{{\em JWST}}

\newcommand\chisq{\ifmmode{\chi^2}\else$\chi^2$\fi}
\newcommand\redchisq{\ifmmode{ \chi\sp{2}\sb{\rm red}}
                    \else\math{\chi\sp{2}\sb{\rm red}}\fi}
\newcommand\Teq{\ifmmode{T\sb{\rm eq}}\else$T$\sb{eq}\fi}
\newcommand\Tb{\ifmmode{T\sb{\rm b}}\else$T$\sb{b}\fi}
\newcommand\mjup{\ifmmode{M\sb{\rm Jup}}\else$M$\sb{Jup}\fi}
\newcommand\rjup{\ifmmode{R\sb{\rm Jup}}\else$R$\sb{Jup}\fi}
\newcommand\msun{\ifmmode{M\sb{\odot}}\else$M\sb{\odot}$\fi}
\newcommand\rsun{\ifmmode{R\sb{\odot}}\else$R\sb{\odot}$\fi}
\newcommand\rstar{\ifmmode{R\sb{\star}}\else$R\sb{\star}$\fi}
\newcommand\fstar{\ifmmode{F\sb{\star}}\else$F\sb{\star}$\fi}
\newcommand\mearth{\ifmmode{M\sb{\oplus}}\else$M\sb{\oplus}$\fi}
\newcommand\rearth{\ifmmode{R\sb{\oplus}}\else$R\sb{\oplus}$\fi}
\newcommand\rplanet{\ifmmode{R\sb{\rm p}}\else$R\sb{\rm p}$\fi}

\newcommand\repack{\textsc{repack}}
\newcommand\pyratbay{\textsc{Pyrat Bay}}
\newcommand\tea{\textsc{tea}}
\newcommand\ggchem{\textsc{GGchem}}
\newcommand\fastchem{\textsc{FastChem}}
\newcommand\mcc{\textsc{mc3}}
\newcommand\chemcat{\textsc{chemcat}}

\newcommand\helios{\textsc{HELIOS}}
\newcommand\multinest{\textsc{MultiNest}}

\newcommand\pcloud{\ifmmode{p\sb{\rm cloud}}
                 \else\math{p\sb{\rm cloud}}\fi}

\newcommand\der{\ifmmode{\rm d}\else\math{\rm d}\fi}

\newcommand\sixtynineb{WASP-69\,b}

\usepackage[bottom]{footmisc}
\usepackage[flushleft]{threeparttable}

\title[Pyrat Bay 2.0: Exoplanet Modeling for JWST]{
Pyrat Bay 2.0: an Upgraded Framework for Exoplanet
Atmosphere Modeling in the JWST Era
}

\author[Patricio E. Cubillos et al.]{Patricio E. Cubillos$^{1,2}$\thanks{E-mail: patricio.cubillos@oeaw.ac.at},
Jasmina Blecic$^{3,4}$,
Denis Shulyak$^{5}$, and
Luca Fossati$^{1}$ \\
$^{1}$Space Research Institute, Austrian Academy of Sciences,
    Schmiedlstrasse 6, A-8042, Graz, Austria \\
$^{2}$INAF -- Osservatorio Astrofisico di Torino,
    Via Osservatorio 20, 10025 Pino Torinese, Italy \\
$^{3}$Department of Physics, New York University Abu Dhabi,
    PO Box 129188, Abu Dhabi, UAE \\
$^{4}$Center for Astrophysics and Space Science (CASS),
    New York University Abu Dhabi, PO Box 129188, Abu Dhabi, UAE \\
$^{5}$Instituto de Astrof\'{\i}sica de Andaluc\'{\i}a - CSIC,
    c/ Glorieta de la Astronom\'{\i}a s/n, 18008 Granada, Spain
}

\date{Accepted XXX. Received YYY; in original form ZZZ}

\pubyear{2026}

\begin{document}
\label{firstpage}
\pagerange{\pageref{firstpage}--\pageref{lastpage}}
\maketitle

\begin{abstract}
This article presents a major update to the open-source {\pyratbay}
modeling framework (version 2.0), tailored for the characterization of
exoplanet atmospheres with the observational capabilities of current
facilities, such as the James Webb Space Telescope ({\JWST}), and
future missions.  The upgraded framework introduces a standalone
chemistry package, {\chemcat}, enabling the modeling of exoplanet
atmospheres via thermochemical-equilibrium calculations with custom
compositions, self-consistent radiative-equilibrium thermal
profiles, and equilibrium-chemistry retrievals.  Other key
implementations include nested sampling via MultiNest, isotopic-ratio
fitting, transit light source modeling, and free-chemistry parameterization allowing vertical
variations in volume mixing ratios.  We validated the
thermochemical- and radiative-equilibrium modules via comparisons with
the {\ggchem}, {\fastchem}, and {\helios} codes.
We present an application of the new modeling framework by conducting
an atmospheric characterization study using simulated transmission
spectra of high signal-to-noise observations for a gas-giant
exoplanet.  We found that vertical abundance variations in planetary
atmospheres can be accurately recovered using data of {\JWST} quality.
In such cases, retrievals that neglect these variations may
result in biased abundance constraints, e.g., when employing the
commonly used free-chemistry approach assuming constant abundance
profiles.
The results highlight the potential of {\JWST} to study
the multi-dimensional nature of exoplanetary atmospheres and their
underlying physical processes.

\end{abstract}



\section{Introduction}

The more than 6000 exoplanets known to date provide a vast and diverse
sample of environments to study planetary physics
\citep[e.g.,][]{Borucki2016rpphKeplerMissionOverview,
  SingEtal2016natHotJupiterTransmission,
  FultonPertigura2018ajGaiaCKS}.  By testing our theoretical models
against observations across this population, we can refine our
understanding of the physical processes that shape both individual
planets and the population as a whole.  This is challenging, however,
because the observed properties of exoplanets reflect the combined
effects of multiple past and ongoing processes throughout their
histories.  The formation path of planets from their protoplanetary
disks determines the building blocks that make up the planets'
composition \citep[e.g.,][]{ObergEtal2011apjCOsnowlines,
  MordasiniEtal2016apjExoplanetFormation,
  PacettiEtal2022apjDiskPlanetsChemicalDiversity}.  Then, hydrodynamic
escape and migration processes can alter the metal fraction of atmospheres over time or
lead to the total loss of primary atmospheres
\citep[e.g.,][]{LammerEtal2003apjEUVatmosphericLoss,
  KubyshkinaEtal2019apjAtmosphericEvolution,
  LoucaEtal2023apjWASP39bChemicalEvolution}.  Lastly, chemical
\citep[e.g.,][]{FegleyLodders1996apjEquilibirumGJ229B,
  VisscherEtal2006apjSPsubstellarChemistry,
  Moses2014rsptaChemicalKinetics} and dynamical processes
\citep[e.g.,][]{ShowmanEtal2009apjRadGCM, AgundezEtal2012aaChemistry}
shape the atmospheric composition of planets across different pressure
levels and latitudinal and longitudinal locations.

Time-series observations of transiting exoplanets has become one of
the most effective tools for probing exoplanet atmospheric
compositions and other properties
\citep{CharbonneauEtal2000apjHD209458bTransit}.  However, interpreting
exoplanet observations is a degenerate problem.  Only through the
detection of multiple species and the precise measurement of their
abundances can we begin to disentangle the impact of the multiple
physical processes mentioned above
\citep[e.g.,][]{FortneyEtal2020ajBeyondEquilibriumTemperature,
  BaeyensEtal2021mnras2DchemistryHorizontalMixing,
  BaeyensEtal2022mnras2Dphotochemistry}.
To constrain multiple species, one needs to distinguish their
unique individual spectral signatures, which requires probing a broad
spectral range at sufficient resolution and signal-to-noise ratio.
Fortunately, the start of operations of new instrumentation has
recently begun to enable such constraints using high-resolution
ground-based observatories
\citep[e.g.,][]{GiacobbeEtal2021natSixMoleculesCNO,
  MansfieldEtal2024ajWASP76bIGRINS} and space-based telescopes.
The {\em James Webb Space Telescope} ({\JWST}) in particular has
demonstrated an unrivaled data quality and the ability to sample a
broad spectral range---potentially observing continuously from
$\sim$0.6 to 12.0~{\microns}---by combining multi-epoch observations
using different instruments
\citep[e.g.,][]{ERSteam2023natCarbonDioxideWASP39b,
  AhrerEtal2023naturJWSTersWASP39bNIRcam,
  AldersonEtal2023naturJWSTersWASP39bG395H,
  CoulombeEtal2023naturWASP18bEmissionNIRISS,
  FeinsteinEtal2023naturJWSTersWASP39bNIRISS,
  RustamkulovEtal2023naturJWSTersWASP39bPRISM}. 
The enhanced characterization capabilities of {\JWST} have
significantly raised the bar for theoretical models, for example, by
requiring a more careful consideration of stellar variability effects
\citep[e.g.,][]{RackhamEtal2018apjStellarHeterogeneityI,
  MoranEtal2023apjGJ486bJWSTwaterOrStar}, planetary spatial
inhomogeneities
\citep[e.g.,][]{ChangeatEtal2019apjTwoLayerParameterization,
  MurphyEtal2024natasLimbAsymmetryWASP107b}, disequilibrium chemistry
processes \citep[e.g.,][]{TsaiEtal2023natWASP39bPhotochemistry}, or
biases arising from the combination of multi-epoch observations
\citep[e.g.,][]{CarterEtal2024natasDataSynthesisWASP39b}.

Generally speaking, there are two primary approaches to inferring the
physical processes behind a given observation: forward modeling and
retrieval techniques.  The forward-model approach relies on well
established physical principles, enabling the simulation of
theoretical scenarios in a self-consistent and highly detailed manner
\citep[e.g.,][]{MarleyEtal1996sciAtmosphereGliese229B,
  MosesEtal2011apjDissequilibriumHD209nHD189b,
  MorleyEtal2015apjFlatSpectra, FossatiEtal2025aaWASP178bNLTE}.
Conversely, retrieval techniques adopt a data-driven approach, using
Bayesian statistical frameworks to obtain statistically robust
constraints of physical parameters
\citep[e.g.,][]{MolliereEtal2019aaPetitRADTRANS, MinEtal2020aaARCiS,
  ZhangEtal2020apjPLATON2, AlRefaieEtal2021apjTaurex3,
  CubillosBlecic2021mnrasPyratBay, WelbanksMadhusudhan2021apjAurora,
  MacDonald2023jossPoseidon, DekaEtal2026apjNEXOTRANS}.  These
statistical frameworks involve computationally intensive processes,
requiring thousands of model evaluations to properly sample the
parameter space \citep{FerozEtal2009mnrasMultiNest,
  ForemanMackeyEtal2013emcee}.  To overcome high computational
demands, retrievals often relax the physical self-consistency of the
models in exchange for performance.  Together, these methods provide a
balanced strategy for understanding exoplanetary atmospheres, with
forward models offering predictive insights and retrievals placing
statistically rigorous estimations based on the observations.

In this context, we present a major update to the open-source
{\pyratbay} framework for the characterization of exoplanet
atmospheres \citep{CubillosBlecic2021mnrasPyratBay}.  Specifically, we
have upgraded the {\pyratbay}'s chemical-equilibrium module, enabling
1D self-consistent radiative-thermochemical-equilibrium calculations
and chemically consistent Bayesian posterior sampling.
We implemented a new modular free-chemistry abundance parameterization
that allows for non-isobaric volume mixing ratios, bridging the gap
between the overtly simplistic constant-with-altitude assumption and
the chemical-equilibrium assumption.
We also implemented parametric models to correct for depth offsets and
scale the uncertainties for combined multi-epoch observations.
Additionally, we added new capabilities such as posterior sampling
with nested sampling, isotopic-ratio modeling, and \ch{H-} absorption,
putting {\pyratbay} up to date with the needs required for exoplanet
atmospheric modeling with {\JWST} and future observatories.

Section \ref{sec:methods} describes and benchmarks the new modules
implemented in the {\pyratbay} framework.  Section
\ref{sec:applications} presents an application on exoplanet
atmospheric characterization using simulated {\JWST} observations.
Section \ref{sec:discussion} summarizes our conclusions.

\section{The {\pyratbay} modeling framework}
\label{sec:methods}

The open-source {\pyratbay} modeling framework
\citep[][]{CubillosBlecic2021mnrasPyratBay} is a modular package that
provides tools for opacity line sampling, planetary atmospheric
modeling, spectral synthesis of transmission and emission
observations, and Bayesian spectral retrieval of exoplanet atmospheres
constrained by transit, secondary-eclipse, or phase-curve
observations.  The following sections describe the most
significant improvements implemented into our framework {\pyratbay}
version 2.0, available at \href{https://pyratbay.readthedocs.io}
{https://pyratbay.readthedocs.io}.

\subsection{Thermochemical equilibrium}
\label{sec:chemcat}

Until now,
{\pyratbay} used the thermochemical-equilibrium
abundances code \citep[{\tea};][]{BlecicEtal2016apsjTEA} to compute
atmospheric abundances.  {\tea} has the flexibility to compute volume
mixing ratios (VMRs) over a wide range of temperatures and pressures
expected for exoplanet atmospheres, but its limited speed makes it less
suitable for computationally intensive applications, such as iterative
forward modeling
or retrieval analyses.  Here we present
Chemistry Calculator for Atmospheres ({\chemcat}), a standalone,
open-source package that builds on our experience with thermochemical
frameworks \citep{BlecicEtal2016apsjTEA, CubillosEtal2019apjRate} and
from other literature packages
\citep[e.g.,][]{AgundezEtal2012aaChemistry,
 StockEtal2018mnrasFastChem, WoitkeEtal2018aaGGchem}.

\subsubsection{Efficient Gibbs free-energy solver}

{\chemcat} computes gas-phase VMRs in thermochemical equilibrium via
Gibbs free energy minimization at known temperatures and pressures,
for a user-defined system of species \citep[as
  in][]{BlecicEtal2016apsjTEA}.  Following
\citet{AgundezEtal2012aaChemistry}, {\chemcat} implements a
Newton-Raphson optimizer with Lagrange multipliers written in C
\citep[see, e.g.,][]{PressEtal2007NumericalRecipes} to impose the
mass-balance constraints as described in
\citet{WhiteEtal1958jcpChemicalEquilibrium}.  This implementation
enables {\chemcat} to compute atmospheric abundances for an atmosphere
in fractions of a second, making the code suitable for iterative or
retrieval applications (e.g., a 100-layer atmosphere is processed in
$\sim$0.02~s with a single CPU on an Intel Core i7-4790 3.60GHz
processor).

In addition, {\chemcat} implements the charge-balance constraint,
assuming a net-neutral mixture, thereby enabling the code to handle
ionic species in the chemical network.  This feature is particularly
relevant for calculations at the high temperatures observed in
planetary atmospheres of ultra hot Jupiters ($T \gtrsim 2000$~K),
where thermal ionization is expected to occur
\citep[e.g.,][]{FossatiEtal2021aaKELT9bNonLTEspectra,
  FossatiEtal2023aaKELT20bBalmerNonLTE}.

\subsubsection{Thermodynamic databases}

As with the predecessor {\tea} code, {\chemcat} uses the NIST-JANAF
database of thermochemical properties \citep{Chase1986jttJANAF}.  This
is an extensive database containing data for over 1700 liquid, solid,
and gaseous molecular species.  It provides tabulated thermochemical
properties, typically (but not always) spanning the 100--6000~K
temperature range, sampled at 100~K intervals. {\chemcat} incorporates
data for 909 neutral and ionic gaseous species from JANAF.

Additionally, {\chemcat} also uses the NASA Glenn
thermodynamic database \citep[Thermo
  Build,][]{McBrideEtal2002nasaThermoBuild}, which contains data for
over 2000 solid, liquid, and gaseous chemical species across a
temperature range of 200 to 20000~K, and also provides the data for
the widely used Chemical Equilibrium with Applications code
\citep[CEA,][]{GordonMcBride1994nasaCEA}.  The Thermo Build data are
expressed as coefficients of a seven-term polynomial for molar heat
capacity at constant pressure and temperature, along with two
integration constants to compute enthalpy and entropy, from which
Gibbs free energy can be derived. {\chemcat} incorporates data for 365
neutral and ionic gaseous species from Thermo Build.

{\chemcat} allows one to select which thermodynamic database to use,
to combine them, or to use custom-provided thermochemical data.  This
flexibility is particularly beneficial for assessing whether the
choice of thermochemical database affects a given application.
In future implementations we will explore the posibility to
  incorporate additional thermodynamic databases such as BURCAT
  \citep{BurcatRuscic2005technoteBURCAT}.

\begin{figure*}
\includegraphics[width=0.85\linewidth,clip]{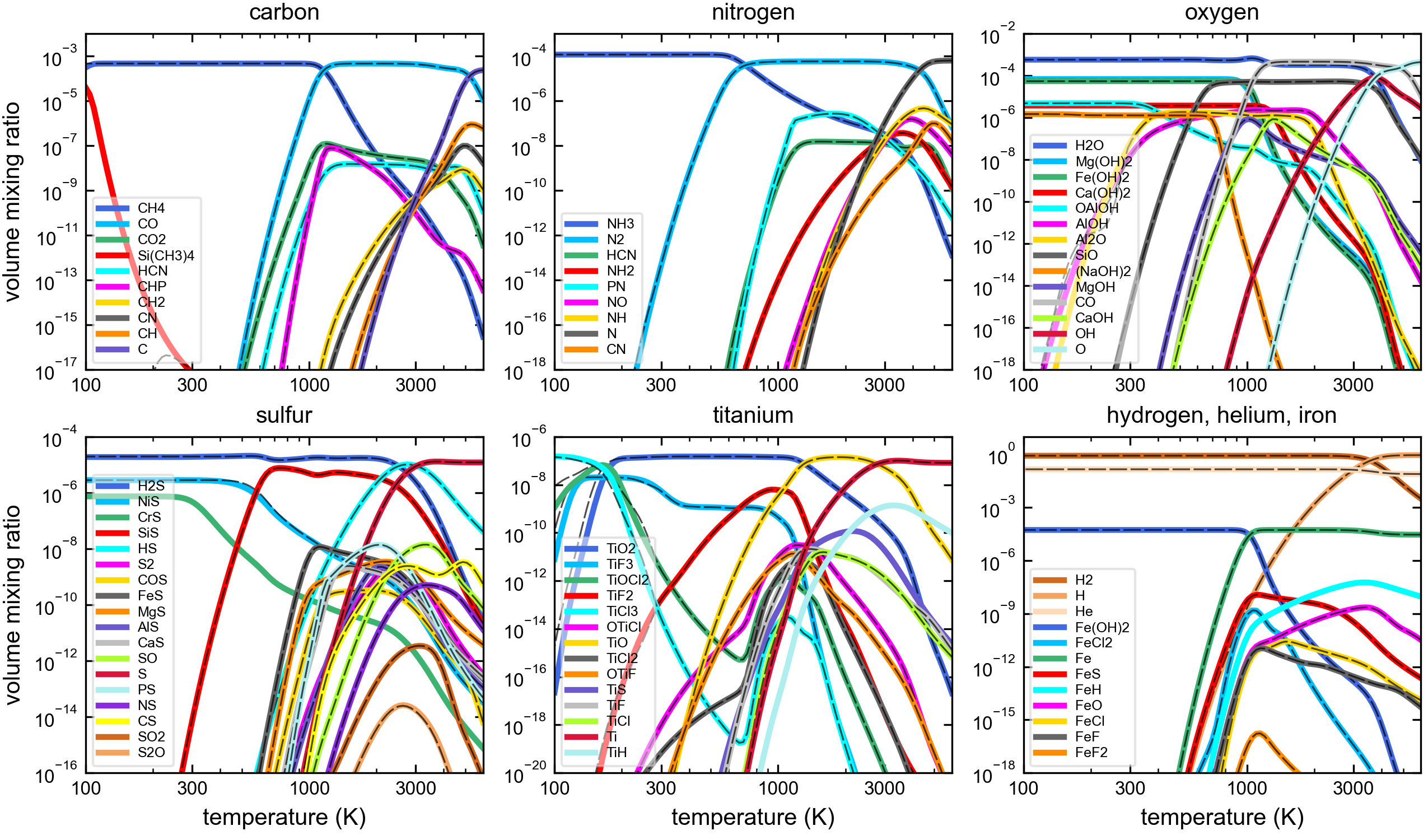}
\caption{Benchmark test between {\chemcat} (dashed black lines) and
  {\ggchem} (solid colored lines) for a network of neutral species at
  1 bar and temperatures from 100--6000~K (only selected species are
  shown).  Both codes produced results typically consistent within 0.1
  dex, with noticeable differences resulting from missing
  thermochemical data or at the lowest temperatures $T \lesssim
  200$~K.}
\label{fig:benchmark_ggchem}
\end{figure*}

\subsubsection{User interface}
\label{sec:chemcat_interface}

The {\chemcat} package offers an intuitive, object-oriented Python
interface, supported by comprehensive unit tests and
documentation\footnote{
\href{https://chemcat.readthedocs.io/en/main/chemistry\_tutorial.html}
     {https://chemcat.readthedocs.io/en/main/chemistry\_tutorial.html}}.
Users initialize a chemical-network object as a one-dimensional
temperature--pressure profile $T(p)$ and a set of atomic and molecular
species.  Since the code is designed for planetary atmospheres,
elemental abundances are initially set to the solar composition
\citep{AsplundEtal2009araSolarComposition, AsplundEtal2021aaTheSun}.
These elemental abundances can be readily customized in multiple ways:

\begin{enumerate}

\item[1.] Use a global 'metallicity' scale factor to:

\begin{enumerate}
\item Scale abundance of all elements other than H and He relative to solar values, [M/H]
\end{enumerate}

\item [2.] Use element-specific abundance scale factors to:

\begin{enumerate}
\item Scale relative to solar values, e.g., [C/H], [O/H], [Na/H]
\item Scale relative to another element, e.g., C/O, K/Na
\end{enumerate}

\end{enumerate}

These parameterization schemes are commonly used in exoplanet
atmospheric retrievals.  For a given scenario, any combination and
number of scaling-factor parameters can be applied as needed.  We note
that element scaling factors relative to solar (2.a) overwrite the
values set by the metallicity scaling factor. Thus, a model
parameterized with [C/H], [O/H], and [M/H] will independently scale
the abundances of carbon, oxygen, and all other metals.
In contrast, element scaling factors relative to other species (2.b)
are applied on top of the other scaling factors. Thus, a model
parameterized with C/O and [M/H] will first scale the metallicity of
all metals and then independently adjust the abundance of carbon
relative to oxygen.

The object-oriented design of {\chemcat} ensures flexibility and
promotes discoverability \citep{MackamulEtal2024hciDiscoverability},
giving users direct access to species' physical properties, such as
equilibrium compositions, stoichiometric values, heat capacity, and
Gibbs free energy.
The {\chemcat} package is available via the Python
Package Index and Conda-forge repositories, simplifying the installation
and integration with other projects, e.g., via the command: \\
\indent \texttt{pip install chemcat} \\
or: \\
\indent \texttt{conda install -c conda-forge chemcat} \\
The source code is available at Github\footnote{\href{https://github.com/AtmoLib/chemcat} {https://github.com/AtmoLib/chemcat}}.

\subsubsection{Equilibrium chemistry benchmarking}

To benchmark the {\chemcat} code, we compared its results with those
of {\ggchem} \citep{WoitkeEtal2018aaGGchem}, a publicly available code
for determining the chemical composition of gases in thermochemical
equilibrium down to 100 K, with or without equilibrium condensation.
We tested both codes on a task similar to that described in Section
5.1 of \citet{WoitkeEtal2018aaGGchem}, involving a chemical network of
442 neutral species derived from 24 different elements: H, He, Li, C,
N, O, F, Na, Mg, Al, Si, P, S, Cl, K, Ca, Ti, V, Cr, Mn, Fe, Ni, Zr,
and W. Both codes calculated thermochemical equilibrium abundances at
1~bar and from 100 to 6000~K. As \citet{WoitkeEtal2018aaGGchem} noted,
the JANAF database does not include many species in the {\ggchem}
network, therefore we opted to include in {\chemcat} only the 183
significant species depicted in Fig.~3 of
\citet{WoitkeEtal2018aaGGchem}.

Figure \ref{fig:benchmark_ggchem} shows a subset of our benchmark test
results.  For most species, we found no significant VMR deviations
between our results (dashed black curves) and {\ggchem} (solid colored
curves).  Notable deviations can be attributed to differences in
extrapolation methods beyond available thermodynamic data (e.g.,
Si(CH$_3$)$_4$ at $T < 300$~K), differences in thermodynamic data
sources (e.g., {\chemcat} uses Thermo Build for SiH$_3$ and SiH$_2$),
missing dominant species (e.g., several Mn molecules), or
the high numerical precision required at the lowest temperatures.
As discussed by \citet{WoitkeEtal2018aaGGchem}, at sufficiently low
temperatures ($T \lesssim 200$~K), some species reach abundances so
low that standard double-precision floating-point representation
becomes inadequate (${\rm VMR} \lesssim 10^{-308}$).  In our
implementation, we opted to cap the values in the Newton-Raphson
routine to avoid VMRs below the minimum representable floating-point
number.
Despite this different approach, we obtained stable results with no
discontinuities and in good agreement between {\chemcat} and
{\ggchem}.  While this method does not provide precise values for the
least abundant species, their low abundances render this discrepancy
negligible for practical applications.
At the lowest temperatures ($T < 200$~K), we found VMR deviations less
than one dex for species in the $10^{-20} < {\rm VMR} < 10^{-8}$
range, and less than 0.25 dex for the more abundant species with ${\rm
  VMR} > 10^{-8}$.  At higher temperatures above 200~K, most VMRs
deviations remain well below the $\sim$0.1 dex level.

\begin{figure}
\includegraphics[width=\linewidth,clip]{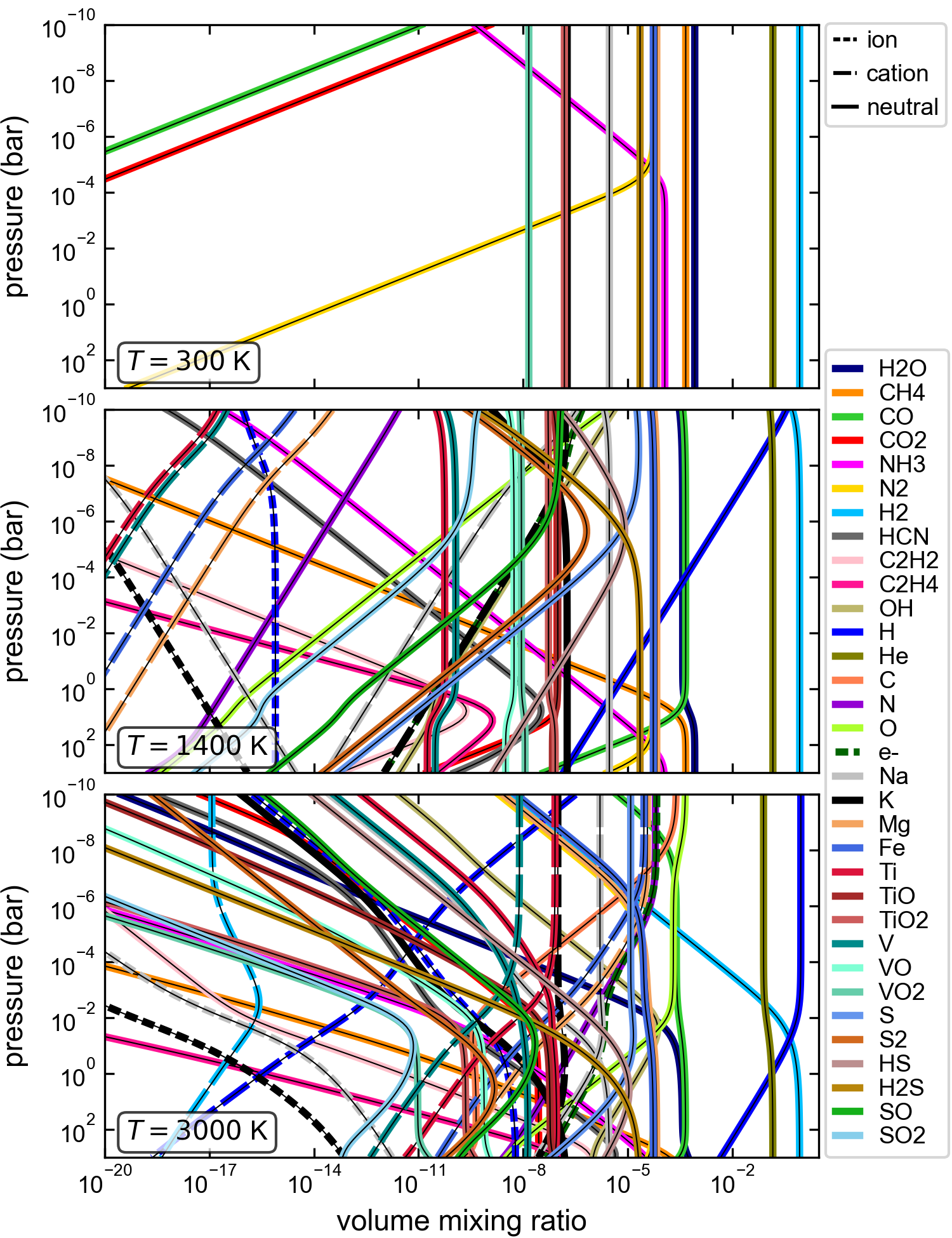}
\caption{Benchmark test between {\chemcat} (thin black lines) and
  {\fastchem} (thick colored lines) for a network of neutral and ionic
  species (see legends), over a $10^3-10^{-10}$~bar range, and at
  three isothermal temperatures of 300, 1400, and 3000~K (top to
  bottom panels, respectively).  Both codes produced results
  consistent to better than 0.05 dex.}
\label{fig:benchmark_fastchem}
\end{figure}

For a second benchmark, we compared {\chemcat} with {\fastchem}
\citep{StockEtal2018mnrasFastChem}, an open-source semi-analytical
program for calculating neutral and ionized gas-phase chemical
equilibria.  Both codes were tested using an identical network of 39
ionic and neutral species, over a temperature range of 300--3000~K and
a pressure range of $10^3-10^{-10}$~bar.
This setup was specifically designed for gas-giant exoplanet
applications, focusing on a chemical network comprising the dominant
species expected for hydrogen, helium, carbon, nitrogen, and oxygen
(\ch{H2O}, \ch{CH4}, CO, \ch{CO2}, \ch{NH3}, \ch{N2}, \ch{H2}, HCN,
\ch{C2H2}, \ch{C2H4}, OH, H, He, C, N, O), their ions (\ch{e-},
\ch{H-}, \ch{H+}, \ch{H2+}, \ch{He+}), alkali metals (Na, \ch{Na-},
\ch{Na+}, K, \ch{K-}, \ch{K+}), and other metals of relevance to
planetary atmospheres (Mg, \ch{Mg+}, Fe, \ch{Fe+}, Ti, TiO, \ch{TiO2},
\ch{Ti+}, V, VO, \ch{VO2}, \ch{V+}).  Figure
\ref{fig:benchmark_fastchem} shows the results of the benchmark test.
Because the same chemical network and thermochemical data (JANAF) were
used by both codes in this comparison, we found an excellent agreement
between the two codes. The abundances for all ionic and neutral
species matched to within 0.05 dex across the entire range of
temperatures and pressures.

\subsection{Radiative equilibrium}
\label{sec:radeq}

Radiative equilibrium occurs when the total bolometric energy entering
a given atmospheric volume equals the energy leaving it. By assuming
radiative equilibrium in a 1D planetary atmospheric model, we can
calculate its temperature profile.

To calculate the energy flowing through each atmospheric layer, our
radiative-equilibrium implementation solves the radiative-transfer
equation, using the two-stream approximation, following the derivation
in \citet{HengEtal2014apjsTwoStreamRT}. Under the assumptions of pure
absorption, hemispheric isotropy (i.e., the intensities are
independent of the zenith angle), local thermodynamic equilibrium, and
a linear dependence of the Planck function $B$ with optical depth
$\tau$, the radiative-transfer solution for the downward and upward
fluxes (denoted as $_\downarrow$ and $_\uparrow$, respectively)
between a pair of upper and lower layers ($u$ and $l$, respectively)
is:
\begin{eqnarray}
\nonumber
F_{\downarrow l}(\lambda) &=& F_{\downarrow u} {\cal T} + \pi B_u (1 - {\cal T}) \\
\label{eq:fdown}
&& +\ \pi B^\prime \left[ -\frac{2}{3} \left(1 - e^{-\Delta\tau}\right)
          + \Delta \tau \left(1 - \frac{{\cal T}}{3}\right) \right], \\
\nonumber
F_{\uparrow u}(\lambda) & = & F_{\uparrow l} {\cal T} + \pi B_l (1 - {\cal T}) \\
\label{eq:fup}
&& +\ \pi B^\prime \left[ \frac{2}{3} \left(1 - e^{-\Delta\tau}\right)
          - \Delta \tau \left(1 - \frac{{\cal T}}{3}\right) \right],
\end{eqnarray}
where $\Delta\tau = \tau_l-\tau_u$, $B^\prime = \Delta B/\Delta\tau$,
and $\cal T$ is the transmission function:
\begin{equation}
{\cal T} = (1-\Delta\tau) e^{-\Delta\tau} + (\Delta\tau)^2 {\cal E}_1(\Delta\tau),
\end{equation}
with ${\cal E}_1$ the exponential integral of the first order.  
Most of these terms are wavelength dependent ($\lambda$),
but we have omitted the $\lambda$ labels for clarity.

To compute fluxes across the atmosphere, we sequentially solve 
Eqs.\ (\ref{eq:fdown}) for $F_{\downarrow l}$ from the top to the
bottom of the atmosphere, and then solve Eqs.\ (\ref{eq:fup}) for
$F_{\uparrow u}$ from the bottom to the top of the atmosphere.  At the
top of the atmosphere, the absorbed stellar irradiation sets the
boundary condition:
\begin{equation}
F_{\downarrow}= \beta_{\rm irr} \left(\frac{\rstar}{a}\right)^2 \fstar,
\end{equation}
where {\rstar} is the stellar radius, {\fstar} is the stellar surface
flux spectrum, $a$ is the semi-major axis, and $\beta_{\rm irr} =
(1-A) f$ is a catch-all term accounting for the Bond albedo $A$
and day--night energy redistribution factor $f$.  For example, values
of $f=1$, 1/2, and 1/4 correspond to absorbed flux at the substellar
point, to uniform redistribution over the day side, and to
redistribution over the entire planet, respectively
\citep[][]{BurrowsEtal2008apjSpectra,
  Guillot2010aaRadiativeEquilibrium}.
\citet{BurrowsEtal2008apjSpectra} further recommends $f=2/3$ for
secondary eclipse observations due to the bias toward the
substellar-point emission \citep[see also Appendix
  in][]{SpiegelBurrows2010apjModelSpectra}.

At the bottom of the atmosphere, the boundary condition is set by an
internal radiative heat term represented by a blackbody with
temperature $T_{\rm int}$, typically set at around 100~K for hot
Jupiters.  In our framework, both $\beta_{\rm irr}$ and $T_{\rm int}$
are adjustable user-defined parameters.

\subsubsection{Radiative and thermochemical equilibrium}

To compute atmospheric profiles in radiative equilibrium, we
implemented the iterative procedure described by
\citet{MalikEtal2017ajHELIOS, MalikEtal2019ajSelfLuminousAtmospheres}.
Using the two-stream radiative-transfer approximation, this approach
integrates the upward and downward fluxes over all wavelengths across
each atmospheric layer to calculate the corresponding bolometric
fluxes: $F_{\uparrow\downarrow}^{\rm bol} = \int
F_{\uparrow\downarrow}(\lambda){\rm d}\lambda$.  From these, the code
determines the net bolometric flux ($F_{\rm net} = F_{\uparrow}^{\rm
  bol} - F_{\downarrow}^{\rm bol}$) and its net divergence ($\Delta
F_{\rm net}$) from the difference in $F_{\rm net}$ between consecutive
layers. The divergence indicates whether an atmospheric layer is in
radiative equilibrium ($\Delta F_{\rm net}=0$), receives more energy
than it emits, or emits more energy than it receives.

To achieve radiative equilibrium, the code iteratively adjusts the
temperature profile so that the net bolometric flux divergence
approaches negligible values at all layers. By combining
Eqs.\ (26--28) into Eq.\ (24) of \citet{MalikEtal2017ajHELIOS}, the
temperature perturbation at iteration $i$ is computed as:
\begin{equation}
\Delta T_i = k_i(p)
    \frac{{\rm sign}(\Delta F_{\rm net})|\Delta F_{\rm net}|^{0.1}}
         {\sigma_{\rm SB} T_i^3\Delta\log p},
\end{equation}
where $\sigma_{\rm SB}$ is the Stefan-Boltzmann constant, $p$ is the
pressure, and $T_i$ is the temperature profile.  $k_i(p)$ is an
adaptive scale factor to modulate the temperature perturbations at
each layer.  $k_i(p)$ is initialized with a value of $10^5$, resulting
in initial perturbations on the order of 100~K.  During each
iteration, $k_i(p)$ is updated, decreasing by a factor of 0.5 at the
layers where the sign of $\Delta F_{\rm net}$ has changed during the
last four iterations, or otherwise increasing by a factor of 1.15.
This scheme effectively damps the magnitude of the perturbations when
the temperature oscillates around a solution \citep[an infamous
  behavior characteristic of radiative equilibrium calculations,
  e.g.,][]{MolliereEtal2015apjPETIT, MalikEtal2017ajHELIOS,
  MukherjeeEtal2023apjPicasso3}.  To prevent unphysical temperature
gradients or ``zig-zag" profiles the code applies a Gaussian kernel
smoothing along the profile.  The kernel's standard deviation is set
to $0.1\times|\Delta T|_{\rm avg}/{\rm K}$, capped between 0.1 and 1.5
dex in pressure, to prevent over-smoothing as the iterative routine
converges toward an equilibrium solution.  At each iteration, the code
invokes the {\chemcat} module to ensure that the abundances remain in
thermochemical equilibrium with the current temperature profile.
Radiative-equilibrium calculations typically converge to a
  stable solution in 100--300 iterations, requiring typically $\sim$10 min of CPU time. For any given
radiative-equilibrium calculation, the metallicity and elemental
composition of the atmosphere can be freely customized as described in
Section \ref{sec:chemcat_interface}.

\subsubsection{Convection}
\label{sec:convection}

Atmospheric energy transport is expected to be dominated by
convection, rather than radiation, when the temperature gradient of
the atmosphere ($\nabla_T = \rm d\log T/\rm d\log p$) becomes steeper
than the adiabatic temperature gradient ($\nabla_{\rm ad}$). This
typically occurs in regions with high atmospheric opacities or low
gravitational acceleration
\citep{CarrollOstlie1996bookModernAstrophysics}. For highly irradiated
exoplanets, convection is generally confined to the deeper atmospheric
layers.

In most atmospheric models, convective energy transport is handled by
identifying convective layers and either enforcing an adiabatic
temperature gradient within them
\citep[e.g.,][]{MolliereEtal2015apjPETIT,
  MukherjeeEtal2023apjPicasso3} or by using mixing-length theory
\citep[e.g.,][]{Rohrmann2001mnrasWhiteDwarfModelAtmospheres,
  GustaffsonEtal2008aaMARCSatmosphericModels,
  GoodyYung1989bookAtmosphericRadiation,
  HenyeyEtal1965apjStellarAtmosphereModels}.
Mixing-length theory is a phenomenological formalism that captures the
general convection behavior observed in stars
\citep[e.g.,][]{CoxGiuli1968bookStellarStructure,
  Mihalas1978bookStellarAtmospheres,
  CarrollOstlie1996bookModernAstrophysics}.  Here we implement a 
convective flux based on the mixing-length formulation of
\citet{CarrollOstlie1996bookModernAstrophysics}:
\begin{equation}
\label{eq:convection}
F_{\rm c} = \alpha^2 \beta^{1/2} \rho c_p T (gH)^{1/2}
    (\nabla_T - \nabla_{\rm ad})^{3/2},
\end{equation}
where $c_p$, $\rho$, $g$, and $H$ are the specific heat capacity at
constant pressure, mass density, gravity, and pressure scale height
throughout the atmosphere, respectively.
The terms $\alpha$ and $\beta$ are arbitrary free parameters of 
mixing length theory:  $\alpha$ quantifies the mixing length traveled
by the convective gas, $l = \alpha  H$, with typical values around
$\alpha\approx1.5$, based on stellar and brown dwarf
atmospheric models \citep{HenyeyEtal1965apjStellarAtmosphereModels,
  Gray1992bookStellarPhotospheres,
  DrummondEtal2016aaChemicalKinetics}.
$\beta$ is a less relevant parameter that accounts for an
approximation of the average speed of the gas, with a value in the
$0<\beta<1$ range
\citep[e.g.,][]{HenyeyEtal1965apjStellarAtmosphereModels,
  CarrollOstlie1996bookModernAstrophysics}.
The convective flux of Eq.\ (\ref{eq:convection}) is added to the
upward bolometric flux at the layers where the atmospheric temperature
gradient is steeper than the adiabatic gradient.

{\renewcommand{\arraystretch}{1.1}
\begin{table}
\centering
\caption{System properties for radiative-equilibrium benchmark. Equilibrium tempertatures were derived from the system parameters assuming zero Bond albedo.}
\label{table:radeq_benchmark}
\begin{tabular*}{1.0\linewidth} {@{\extracolsep{\fill}} lccc}
\hline
System      & WASP-107 &  WASP-39 &  WASP-121  \\
\hline
{\bf Host star:} \\
$T_{\rm eff}$ (K)          & 4400.0 & 5500.0  & 6400.0 \\
$R_{\rm star}$ ($\rsun$)   & 0.67   & 0.94  & 1.458 \\
{\bf Planet:} \\
$R_{\rm planet}$ ($\rjup$) & 0.943  & 1.280  & 1.753 \\
$M_{\rm planet}$ ($\mjup$) & 0.096  & 0.281  & 1.157 \\
semi-major axis (AU)    & 0.055  & 0.048  & 0.026 \\
$T_{\rm eq}$ (K)          & 750.0  & 1200.0  & 2300.0 \\
\hline
\end{tabular*}
\end{table}
}

\subsubsection{Radiative thermochemical equilibrium benchmark}
\label{sec:benchmark}

To validate our radiative-equilibrium scheme, we performed benchmark
simulations, comparing our results with those generated by the
{\helios} package \citep{MalikEtal2017ajHELIOS}. These tests consist
of six scenarios involving three planetary systems representing a
distinct stellar-irradiation regime: a warm Jupiter (similar to
WASP-107\,b), a classical hot Jupiter (WASP-39\,b), and an ultra-hot
Jupiter (WASP-121\,b).
We modeled two scenarios for each planet, first adopting a default
configuration for all of them, i.e., assuming solar composition, no
internal heat flux ($T_{\rm int} = 0$~K), and a common set of the
optical and infrared cross sections.  Then, for each
planet we explored an additional scenario by varying one specific
model parameter: for WASP-107\,b we simulated an atmosphere with a
higher internal heat flux of $T_{\rm int} = 350$~K, for WASP-39\,b a
higher atmospheric metallicities of 50$\times$ solar, and for
WASP-121\,b an atmosphere with additional TiO and VO optical
absorbers.
Table \ref{table:radeq_benchmark} summarizes the system parameters
adopted for these benchmark simulations.

\begin{figure}
\includegraphics[width=\linewidth,clip]{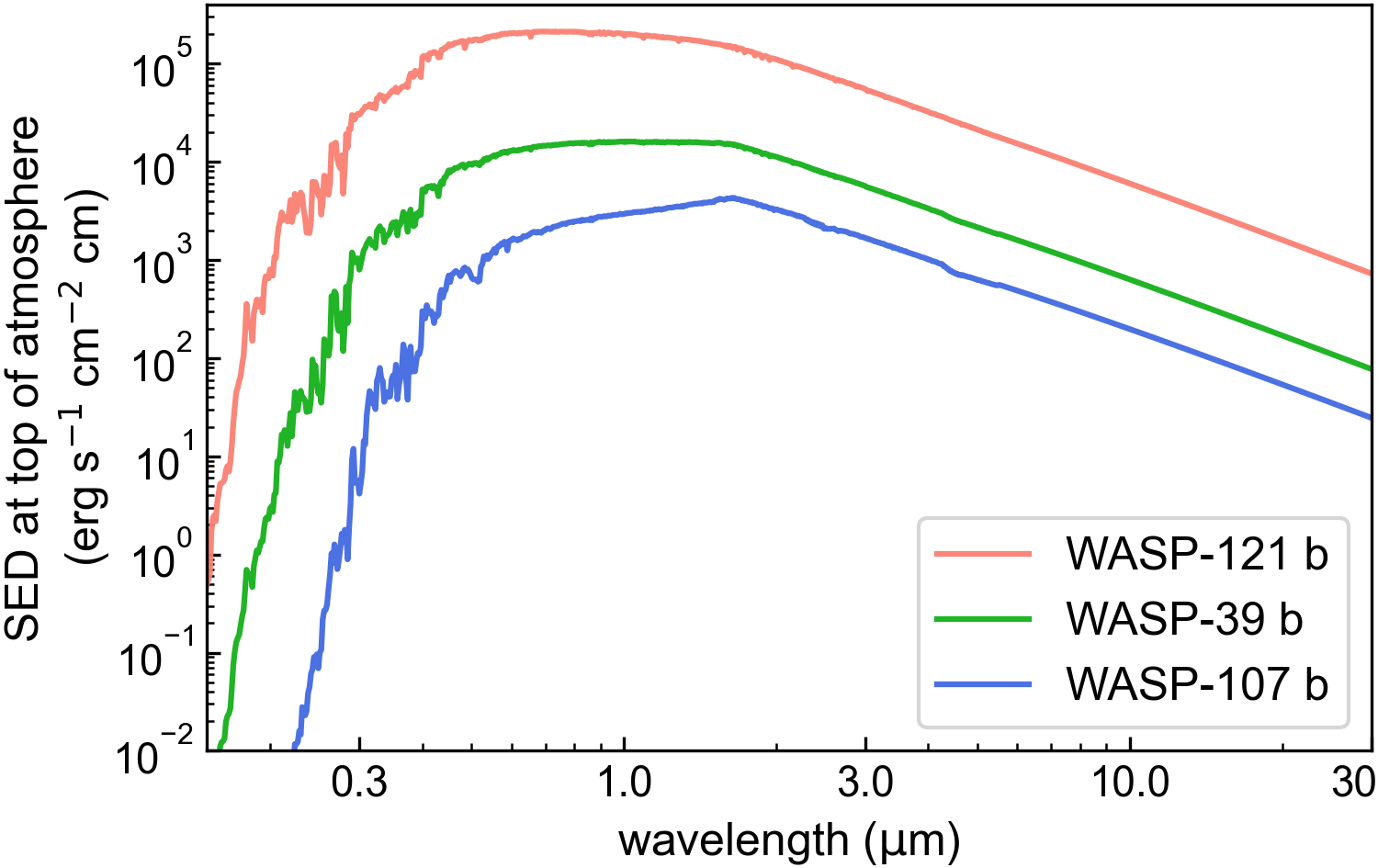}
\caption{Incident stellar SEDs at the top of the atmosphere for the
  radiative-equilibrium benchmark runs (see labels).  These are
  PHOENIX models \citep{HusserEtal2013aaPHOENIXstellarModels} with
  solar metallicity, surface gravity of $\log\ g=4.5$, and effective
  temperatures of 4400~K (WASP-107, spectral type K4V), 5500~K
  (WASP-39, G8V), and 6400~K (WASP-121, F5V).  }
\label{fig:benchmark_sed}
\end{figure}

For the host stars' spectral energy distribution (SEDs) we adopted
PHOENIX models \citep{HusserEtal2013aaPHOENIXstellarModels}, assuming
solar metallicity and a surface gravity of $\log(g)=4.5$
(Fig.\ \ref{fig:benchmark_sed}).

Both {\pyratbay} and {\helios} adopted a similar planetary atmosphere
model, with pressures ranging from 100 to $10^{-9}$ bar, sampled with
81 layers for {\pyratbay} and 115 for {\helios}.  Both codes also
adopted a wavelength spectrum ranging from 0.15 to 33~{\microns}
sampled at a resolution $R = \lambda/\Delta\lambda = 3000$.  This
broad spectral range was chosen to capture the bulk of radiation from
both the host star and planet.  We included the same set of atomic and
molecular cross sections, sampled over this wavelength array.
Their abundances were calculated under thermo-chemical equilibrium using
{\chemcat} (\pyratbay) and \textsc{FastChem} ({\helios}).
For the atomic species we computed custom cross sections for Na, K,
Fe, \ch{Fe+}, Mg, and \ch{Mg+}, using the line lists from the Kurucz
database \citep{Kurucz2018aspcOpacities}.  For the molecular species
we included cross sections obtained from
DACE (\href{https://dace.unige.ch}{dace.unige.ch}),
which were based on
the HITEMP and ExoMol line-lists for \ch{H2O}
\citep{PolyanskyEtal2018mnrasPOKAZATELexomolH2O}, CO
\citep{LiEtal2015apjsCOlineList}, \ch{CO2}
\citep{YurchenkoEtal2020mnrasCO2ucl4000}, \ch{CH4}
\citep{YurchenkoTennyson2014mnrasExomolCH4}, \ch{NH3}
\citep{Yurchenko2015jqsrtBYTe15exomolNH3,
  ColesEtal2019mnrasNH3coyuteExomol}, HCN
\citep{HarrisEtal2006mnrasHCNlineList, HarrisEtal2008mnrasExomolHCN},
\ch{H2S} \citep{AzzamEtal2016mnrasExoMolH2S}, \ch{SO2}
\citep{UnderwoodEtal2016mnrasSO2exoamesExomol}, \ch{C2H2}
\citep{ChubbEtal2020mnrasC2H2acetyExomol}, TiO
\citep{McKemmishEtal2019mnrasTOTOexomolTiO} and VO
\citep{McKemmishEtal2016mnrasVOMYTexomolVO}.
These cross sections are pre-calculated at a fixed grid in
  temperature, pressure, and wavelength.  The radiative transfer
  calculation interpolates from these tables at each
  radiative-equilibrium iteration.
For the continuum cross sections, we included collision-induced
absorption for \ch{H2}--\ch{H2} and \ch{H2}--He
\citep{BorysowEtal1988apjH2HeRT, BorysowEtal1989apjH2HeRVRT,
  BorysowEtal2001jqsrtH2H2highT, BorysowFrommhold1989apjH2HeOvertones,
  Borysow2002jqsrtH2H2lowT}; Rayleigh scattering for \ch{H2}, H, He,
and e$^-$ \citep{Kurucz1970saorsAtlas}; and \ch{H-} free-free and
bound-free absorption \citep{John1988aaHydrogenIonOpacity}.

\begin{figure*}
\includegraphics[width=\linewidth,clip]{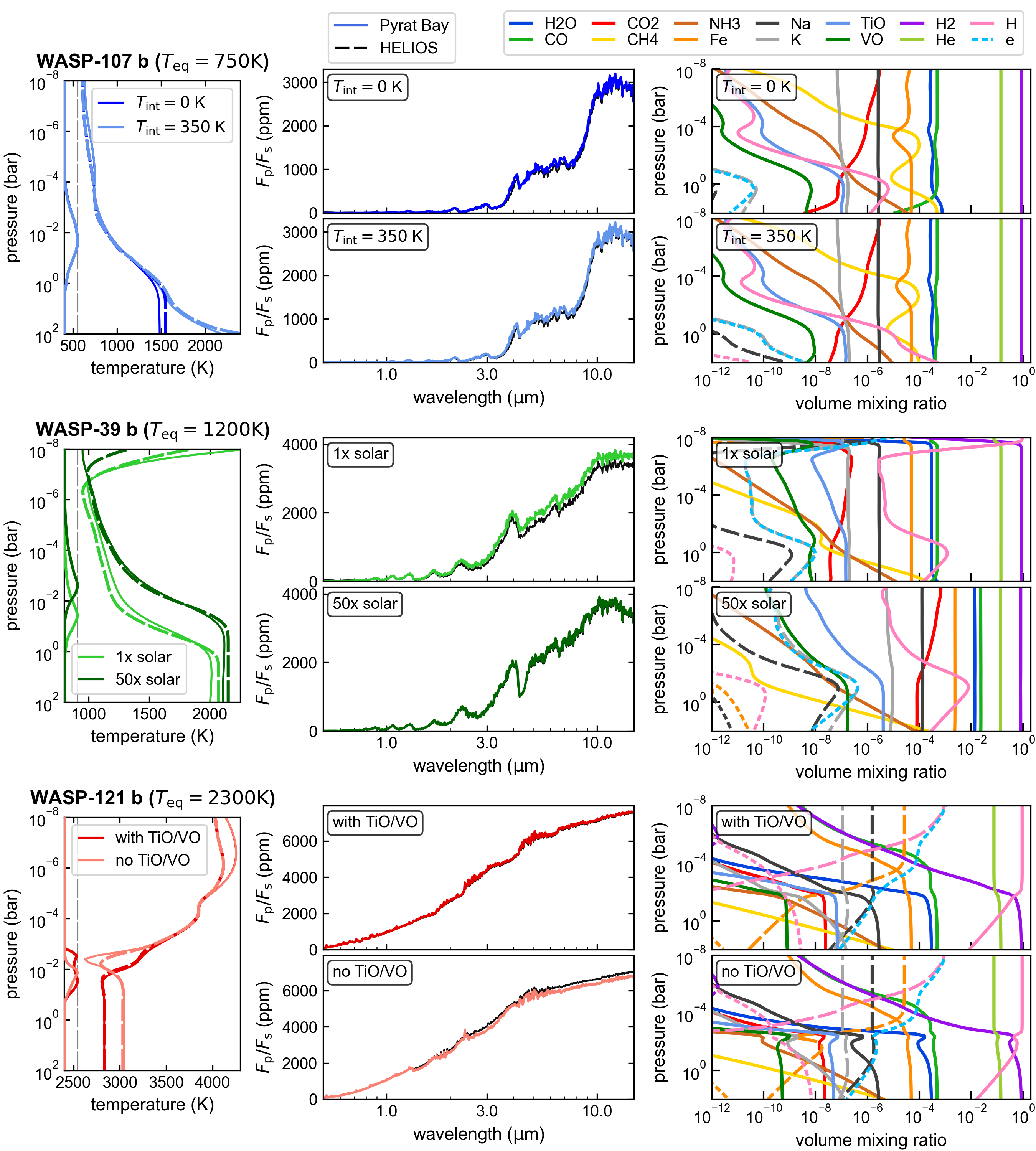}
\caption{Radiative-equilibrium benchmark between {\pyratbay} and
  {\helios}.  The top, middle, and bottom rows show simulations for a
  temperate planet (WASP-107\,b), a hot-Jupiter (WASP-39\,b), and an
  ultra-hot Jupiter (WASP-121\,b), respectively.  {\bf Left column:}
  radiative-equilibrium temperature profiles estimated by {\pyratbay}
  (solid curves) and {\helios} (dashed curves).  Light and dark colors
  correspond to two simulated scenarios for each system (see
  legend). The contribution-function curves on the left of each panel
  show the region of the atmosphere probed by NIR eclipse
  observations. {\bf Center column:} derived planet-to-star flux ratio
  (eclipse spectra) for each simulation, as computed by {\pyratbay}
  (colored) and {\helios} (black).  {\bf Right column:} atmospheric
  volume mixing ratios (VMR) derived by {\pyratbay} for each
  simulation for a selection of the most relevant
  species (see legend on the top right corner).  Neutral species,
  negative ions, and positive ions are shown as solid, short-dashed, and long-dashed
  curves, respectively. The {\helios} VMRs follow the same trends as
  shown by {\pyratbay} (omitted for clarity), being typically
  consistent within $\sim$0.1 dex, where the main discrepancies are
  consistently explained by the difference in temperature.}
\label{fig:benchmark_radeq}
\end{figure*}

Figure \ref{fig:benchmark_radeq} shows the temperature profiles,
emission spectra, and composition of the radiative-equilibrium
benchmark simulations.  Overall we find good agreement between
{\pyratbay} and {\helios} across all scenarios.
The typical temperature-profile differences between {\pyratbay} and
{\helios} are in the 10--50~K range for these scenarios, whereas the
NIR emission spectra differ by 1--10\% on average.
More importantly, the radiative-equilibrium outputs follow the
expected behavior according to the change of environmental conditions.
Increasing the incident stellar irradiation leads to hotter
atmospheres with distinct thermal structures: the least irradiated
case (WASP-107\,b) shows a non-inverted profile; the hot Jupiter
(WASP-39\,b) shows heating confined to the uppermost layers; and the
ultra-hot Jupiter (WASP-121\,b) shows a strong thermal inversion.

The two WASP-107\,b simulations illustrate how a higher internal heat
flux yields warmer deep layers, with the excess heat quickly dissipating 
at higher pressures.
The two WASP-39\,b simulations show how an increased atmospheric
metallicity shifts the photosphere to lower pressures, as the enhanced
metallicity increases the overall opacity of the atmosphere.
Finally, the two WASP-121\,b simulations highlight how optical absorbers
not only drive thermal inversions but also impact the deep atmospheric
layers.  As the stellar radiation dominates the energy budget of
strongly irradiated planets, a higher optical opacity intercepts more
flux at low pressures, reducing the energy that reaches depth and
yielding lower temperatures in the bottom layers.
In this regard, Na and K absorption has a significant role in shaping
the thermal structure for all of these simulations.  For ultra-hot
Jupiters, atomic absorption by Fe, Mg, \ch{Fe+}, and \ch{Mg+}
also have a significant impact shaping the thermal structure of planets.

Considering the large range of parameters that can factor into a
radiative-equilibrium calculation, different results between different
implementations are expected to occur.
A major factor impacting the radiative-equilibrium calculations is the
choice of the opacity source for a given absorber.  Particularly, we
found that the choice of the alkali opacities (e.g.,
\citealp{BurrowsEtal2000apjBDspectra} or
\citealp{AllardEtal2019aaSodiumProfile}) can introduce discrepancies
on the order of $\sim$100~K. Further discrepancies can arise from the
choice of line profiles when sampling line lists (i.e., choice of
line-wing extent or broadening parameters), choice for which there is
no general prescription other than {\em ad-hoc} values adopted by the
community \citep[see, e.g.,][]{GrimmHeng2015apjHELIOSK}.  We also
found that the choice of spectral resolving power (R=3000
  vs. 25000) can lead to differences of dozens of K.

\subsection{Bayesian atmospheric retrievals}

\subsubsection{Nested sampling}
\label{sec:sampling}

Up until now, {\pyratbay} has implemented the Differential-evolution
Markov-chain Monte Carlo (MCMC) Bayesian sampler
\citep{terBraak2008SnookerDEMC} via {\mcc}
\citep{CubillosEtal2017apjRednoise}.  An alternative to MCMC sampling is 
nested sampling
\citep{Skilling2004aipcNestedSampling, Skilling2006baNestedSampling}.
Nested sampling works by evolving a collection of live points that
sample the prior parameter space.  At each iteration, the algorithm
replaces the lower-likelihood point with a new one that has a larger
likelihood.  In this way, this process continuously shrinks the prior
volume, thus converging towards the highest likelihood region(s).
Nested sampling then estimates the Bayesian evidence by integrating
the parameter space, from which the parameter posterior distribution
can be derived.  This is an advantage of nested sampling over MCMC,
since the evidences allow for parameter inference and model
comparison.  For a more in-depth overview of the nested-sampling
algorithm and its application in the physical sciences see
\citet{AshtonEtal2022nrvmpNestedSampling}.
In {\pyratbay} we have incorporated the widely-used {\multinest}
nested-sampling implementation \citep{FerozEtal2009mnrasMultiNest} via
the pyMultiNest Python interface
\citep{BuchnerEtal2014aaBayesianXrayAGN}.  Atmospheric retrieval
employing {\multinest} can be executed in parallel via Message Passing
Interface (MPI).

\subsubsection{Free-chemistry retrievals with non-isobaric VMRs}

The 'free-chemistry' parameterization is a common assumption used in
retrieval studies \citep[e.g.,][]{BellEtal2024natasWASP43bMIRIphase,
  BennekeEtal2024arxivTOI270dMiscible,
  VermaEtal2025ajHD209458btransits}.  In this approach, the abundances
of species are treated as independent constant-with-altitude (i.e.,
isobaric) volume mixing ratios (VMRs).  This is an ad hoc
approximation suitable when thermochemical equilibrium is not expected
(e.g., temperate atmospheres) or when observations lack the
sensitivity to constrain the vertical structure of a VMR in a
statistically significant manner.  However, simulations
\citep{ChangeatEtal2019apjTwoLayerParameterization} and recent
observations \citep[e.g.,][]{GaraiEtal20025aaKELT7bCHEOPS,
  ChallenerEtal2025natasWASP18bSOSSmapping} suggest that more
realistic models may be required, particularly for {\JWST}
observations of gas giants.  For this reason, we have developed a new
flexible non-isobaric VMR model.

\begin{figure}
\includegraphics[width=\linewidth,clip]{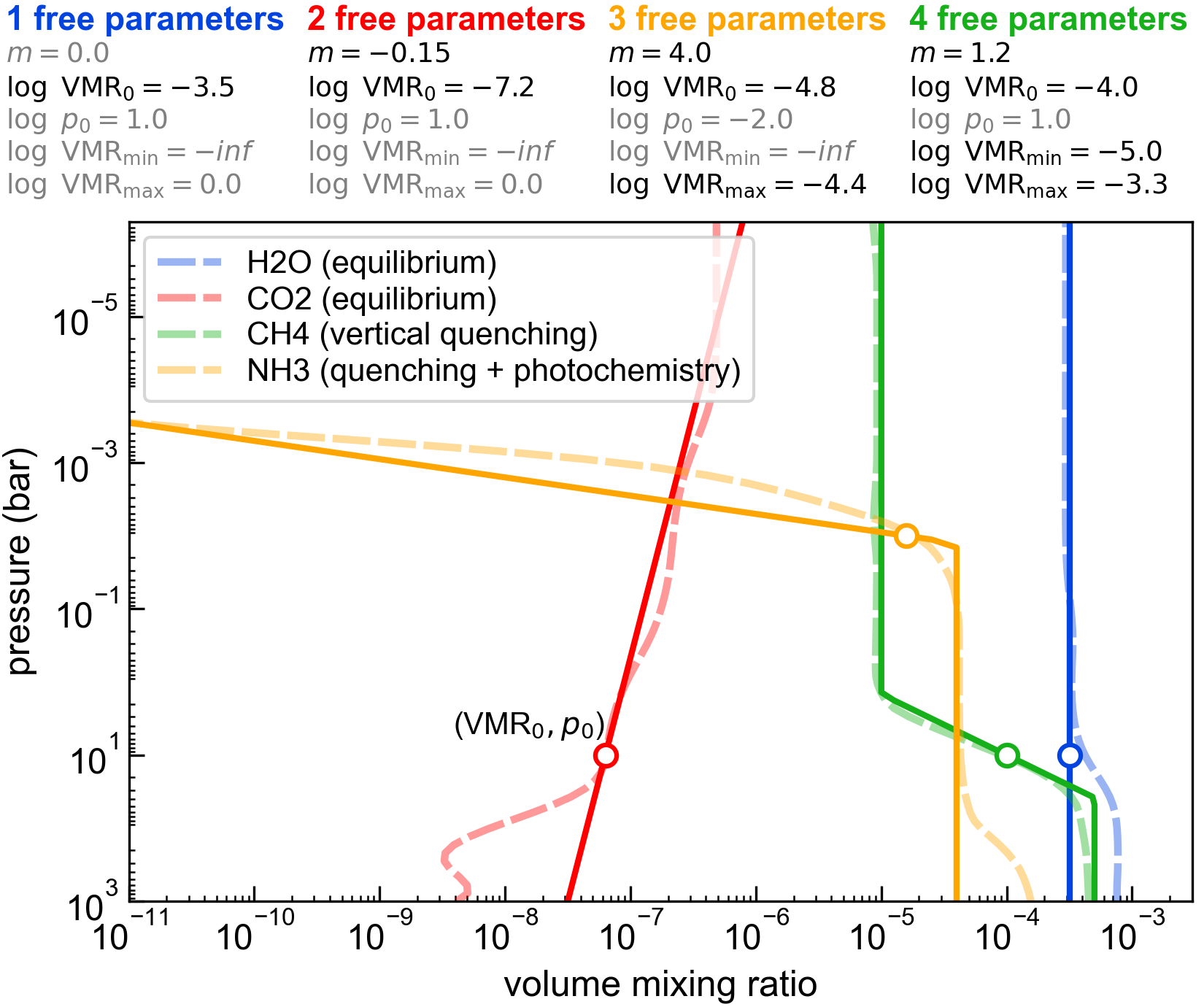}
\caption{Examples of VMR profiles expected in planetary atmospheres
  \citep[dashed curves, adapted
    from][]{MosesEtal2011apjDissequilibriumHD209nHD189b} and
  parametric non-isobaric VMR profiles (solid curves). The
  non-isobaric VMR model can adjust to the complexity of the profile
  by varying the number of free parameters, for example: isobaric
  profiles (blue, one free parameter), slanted profiles (red, two free
  parameter), quenched and photochemically affected profiles (yellow,
  three free parameters), or vertically quenched profiles (green, four
  free parameters).  The text above the panel lists the values for each VMR
  model, with black and gray text indicating free and fixed
  parameters, respectively.}
\label{fig:vmr_model}
\end{figure}

The non-isobaric model consist of a slanted $\log {\rm VMR}$--$\log p$
profile capped between two VMR values.
The model is parameterized by a slope $m$ and reference
  point \{$\log {\rm VMR}_0$, $\log p_0$\} that define a slanted VMR
  curve, and by two additional parameters that cap the profile at a
  minimum and maximum VMR value.
This parameterization allows the model to simulate
a wide variety of VMR profiles expected from equilibrium and
disequilibrium-chemistry calculations (Fig.\ \ref{fig:vmr_model}).
Note that \{$\log$VMR$_0$, $\log p_0$\} are a degenerate pair of
values; for retrieval analyses one parameter should be kept fixed,
while leaving the other one free.  Thus, only between one and four
free parameters are needed for practical applications.  This
highlights the flexibility of the model and its adaptability to the
complexity of the problem at hand (Fig.\ \ref{fig:vmr_model}). In its
simplest form with a single free parameter ($\log$VMR$_0$), the model
reduces to the isobaric case (thus enabling nested-sampling model
comparison with the standard free-chemistry approximation).  Two free
parameters ($m$ and $\log$VMR$_0$) can fit a monotonically decreasing
(or increasing) VMR profile.  Three free parameters ($m$,
$\log$VMR$_0$, and $\log$VMR$_{\rm min}$) can fit profiles that are a
combination of constant and slanted sections. Four free parameters can
fit abundances constrained between two values.  See Section
\ref{sec:applications} for a worked example.

\subsubsection{Chemically-consistent atmospheric retrievals}

The `equilibrium-chemistry' parameterization is a second approach commonly used
 used for exoplanet atmospheric retrievals.  Here, the
composition is parameterized by the abundance of the elemental
building blocks, and computed self-consistently with the
pressure--temperature profile under the assumption of thermochemical
equilibrium.

In {\pyratbay} we implemented the equilibrium-chemistry retrieval
approach by employing the new equilibrium-chemistry module (Section
\ref{sec:chemcat}), ensuring that the model has the flexibility to
vary the abundance of any element in the atmosphere with a variety of
parameterizations.  This is relevant given the rich information
content provided by current observations
\citep[e.g.,][]{PelletierEtal2026aaWASP121bNIRISS}.
As a baseline, the code assumes a solar elemental abundance
composition \citep{AsplundEtal2021aaTheSun}.  Free parameters can
modify elemental abundances as a whole by a metallicity factor [M/H]
that scales all metals relative to solar abundances, or individually:
either relative to the solar abundance (e.g., [C/H], [O/H], etc.) or
relative to other elemental abundances (e.g., C/O, Na/K, etc.). Any
number of these free parameters can be used and combined as needed.

\subsubsection{Hybrid equilibrium plus disequilibrium chemistry}

A third retrieval mode implemented in {\pyratbay} is a hybrid approach
that combines equilibrium-chemistry abundances with additional
free-chemistry VMR parameters for selected species.  This approach is
suitable when neither the free-chemistry nor chemical-equilibrium
assumptions alone can adequately describe an atmospheric composition.
A common example is when modeling atmospheres with out-of-equilibrium
\ch{SO2} abundances, influenced by quenching and photochemistry
processes \citep[e.g.,][]{TsaiEtal2023natWASP39bPhotochemistry,
  KirkEtal2025mnrasWASP15bNIRSpec}.

\subsubsection{Isotopic-ratio fitting}

The measurement of isotopic ratios has recently emerged as a powerful
tool for characterizing sub-stellar objects, their atmospheres, and
their formation paths \citep{MolliereSnellen2019aaDetectingIsotopicRatios}.
This method relies on resolving and measuring individual absorption
lines, typically requiring high signal-to-noise ratio and high
spectral-resolution observations.  For this reason, isotopic ratios
have been detected mainly in directly imaged objects from ground-based
\citep[e.g.,][]{ZhangEtal2021natYSES1bCarbon13,
  GandhiEtal2025mnrasABPicbCRIRESplus} and space telescopes
\citep{GandhiEtal2023apjVHS1256bNIRSpec}.

We enabled in {\pyratbay} the capability to fit and retrieve
custom isotopic ratios via
isotopic-fraction parameters ($\log\delta^{\rm iso} X$), which model
the log-scale isotopic fraction of selected isotopes of a given
species.  These parameters can either take specific values, e.g.,
$\log\delta^{13} {\rm CO} = -1.9$, or act as ``filler'' values (analogous
to free-chemistry VMR fitting), e.g.:
\begin{equation}
\delta^{12}{\rm CO} = 1 - \delta^{13}{\rm CO} - \delta{\rm C}^{18}{\rm O}.
\end{equation}
This provides a fully customizable framework for modeling isotopic
ratios in both forward modeling and retrieval analyses.

Applying isotopic-ratio weighting requires the species' input cross
sections to be partitioned by isotope.  For this, we upgraded the
{\pyratbay} line-sampling routine to generate cross-section
tables for selected isotopes from ExoMol or HITRAN line lists.
Spezzano et al.\ (in prep.) present a validation of this
implementation by applying it on a brown-dwarf observation.

\subsection{Cross sections}

\subsubsection{\ch{H-} cross sections}

\ch{H-} continuous absorption can become significant at the high
temperatures expected for ultra Hot Jupiters \citep[$T\gtrsim
  2000$~K,][]{ArcangeliEtal2018apjWASP18bEmissionHminus}, where
molecular hydrogen dissociates to give way to atomic and ionic
hydrogen as the most abundant species.  We have implemented the
\ch{H-} absorption model from \citet{John1988aaHydrogenIonOpacity},
which accounts for bound-free photo-detachment (\ch{$h\nu$ +
  H-}~$\rightarrow$~\ch{H + e-}) and free-free scattering (\ch{$h\nu$
  + e- + H}~$\rightarrow$~\ch{H + e-}).

The bound-free and free-free cross sections for \ch{H-} (in units of
cm$^{5}$ per hydrogen atom per electron) at a temperature $T$ (in K) and
wavelength $\lambda$ (in {\microns}) are:

\begin{align}
\nonumber
  \sigma_{\rm bf}(T,\lambda) = & 0.75 \times 10^{-18} k_{\rm B} T^{-3/2}
      e^{\alpha/\lambda_0 T} \left(1-e^{-\alpha/\lambda T}\right)\times \\
\label{eq:h_bf}
& \lambda ^3 \left(\frac{1}{\lambda}-\frac{1}{\lambda_0}\right)^{3/2}
      \sum_{n=1}^{6} c_n \left(\frac{1}{\lambda}-\frac{1}{\lambda_0}\right)^{(n-1)/2}, \\
\nonumber
  \sigma_{\rm ff}(T,\lambda) = & 10^{-29} k_{\rm B}T
      \sum_{n=1}^{6} \left(\frac{5040}{T}\right)^{(n+1)/2}\times \\
\label{eq:h_ff}
& \left\{\lambda^2 A_n + B_n + \frac{C_n}{\lambda} + \frac{D_n}{\lambda^2} + \frac{E_n}{\lambda^3} + \frac{F_n}{\lambda^4} \right\},
\end{align}

where $\lambda_0=1.6419$~{\microns} is the photo-detachment
threshold, $k_{\rm B}$ is the
Boltzmann constant, and $\alpha= hc/k_{\rm B} = 1.439$~cm\,K.  The polynomial coefficients in Eqs.\ (\ref{eq:h_bf}) and
(\ref{eq:h_ff}) are listed in \citet{John1988aaHydrogenIonOpacity}.

\subsubsection{Tabulated cross sections}

While {\pyratbay} continues to enable sampling custom cross sections,
we also provide pre-computed cross-section files intended for a broad
range of applications.  These cross sections have been computed
assuming a \ch{H2}/He-dominated atmosphere, terrestrial isotopic
ratios, and Voigt profiles with wing cut-offs at 300 half-width at
half maximum (HWHM) and at 25 cm$^{-1}$.  The grids have been sampled
in wavelength from 0.15 to 33~{\microns} at a resolving power of
$R=25,\!000$, in temperature from 200 to 5000~K with a step of 150~K,
and in log-pressure from $10^{-9}$ to 1000~bar with 4 samples per dex.
We selected a resolution of $R=25,\!000$ following
  \citet{Leconte2021aaExoK} and Welbanks et al.\ (in prep.), which
  found that line-sampling resolutions of $20,\!000-30,\!000$ reach
  transmission-spectrum accuracies of $\sim$$1-10$ ppm, which is at or
  below the typical uncertainty of {\JWST} observations.

The cross-section files are hosted in a Zenodo repository reachable at
\href{https://zenodo.org/records/16965390}{https://zenodo.org/records/16965390}.
At the time of publication we provide cross sections for 38 species
from the HITRAN, Exomol, and Ames databases (Table
\ref{table:cross_sections}).  The largest line lists
  consisting of several billions of transitions have been
  pre-processed with the {\repack} algorithm
  \citep{Cubillos2017apjRepack} to extract the dominant
  transitions. New cross sections will be added or updated as they
  become available.

{
\begin{table}
\centering
\caption{Cross-sections currently provided in {\pyratbay}, available
  at this repository:
  \href{https://zenodo.org/records/16965390}{https://zenodo.org/records/16965390}}
\label{table:cross_sections}
\begin{tabular*}{1.0\linewidth} {@{\extracolsep{\fill}} rcl}
\hline
Species   & Source & References  \\
\hline
\ch{AlF}  & Exomol & \citet{Bernath2020jqsrtMoLLIST}  \\
\ch{C2H2} & Exomol & \citet{ChubbEtal2020mnrasC2H2acetyExomol}  \\
\ch{CH4}  & Exomol & \citet{YurchenkoEtal2024mnrasExomolCH4mm}  \\
\ch{CO}   & HITEMP & \citet{LiEtal2015apjsCOlineList}  \\
\ch{CO2}  & Ames   & \citet{HuangEtal2023jmospAmesCO2}  \\
\ch{CS}   & Exomol & \citet{PauloseEtal2015mnrasExomolCS} \\
\ch{CS2}  & HITRAN & \citet{GordonEtal2026jqsrtHITRAN2024}  \\
\ch{CaH}  & Exomol & \citet{OwensEtal2022mnrasExomolCaH}  \\
\ch{Fe}   & VALD   & \citet{PiskunovEtal1995aapsVALDdatabase}  \\
\ch{FeH}  & Exomol & \citet{DulickEtal2003apjFeHopacity, Bernath2020jqsrtMoLLIST}  \\
\ch{H2O}  & Exomol & \citet{PolyanskyEtal2018mnrasPOKAZATELexomolH2O}  \\
\ch{H2S}  & Exomol & \citet{AzzamEtal2016mnrasExoMolH2S, ChubbEtal2018jqsrtExomolH2S}  \\
\ch{HCN}  & Exomol & \citet{HarrisEtal2008mnrasExomolHCN, BarberEtal2014mnrasExomolHCN}  \\
\ch{HCl}  & HITRAN & \citet{GordonEtal2026jqsrtHITRAN2024}  \\
\ch{HF}   & Exomol & \citet{CoxonHajigeorgiou2015jqsrtHFandHBr}  \\
\ch{K}   & VALD   & \citet{PiskunovEtal1995aapsVALDdatabase}  \\
\ch{KCl}  & Exomol & \citet{BartonEtal2014mnrasExomolNaClKCl}  \\
\ch{KOH}  & Exomol & \citet{OwensEtal2021mnrasExomolKOHandNaOH}  \\
\ch{Mg}   & VALD   & \citet{PiskunovEtal1995aapsVALDdatabase}  \\
\ch{Na}   & VALD   & \citet{PiskunovEtal1995aapsVALDdatabase}  \\
\ch{NaCl} & Exomol & \citet{BartonEtal2014mnrasExomolNaClKCl}  \\
\ch{NaH}  & Exomol & \citet{RivlinEtal2015mnrasExomolNaH}  \\
\ch{NH3}  & Exomol & \citet{ColesEtal2019mnrasNH3coyuteExomol, YurchenkoEtal2024mnrasExomolNH3coyute}  \\
\ch{OCS}  & Exomol & \citet{OwensEtal2024mnrasExomolOCS}  \\
\ch{OH}   & HITEMP & \citet{GordonEtal2022jqsrtHITRAN2020}  \\
\ch{PH}   & Exomol & \citet{LanglebenEtal2019mnrasExomolPH}  \\
\ch{PH3}  & Exomol & \citet{SousaSilvaEtal2015mnrasExomolPH3}  \\
\ch{PN}   & Exomol & \citet{SemenovEtal2025mnrasExomolPN}  \\
\ch{PO}   & Exomol & \citet{PrajapatEtal2017mnrasExomolPOPS}  \\
\ch{SH}   & Exomol & \citet{GormanEtal2019mnrasGYTexomolSH}  \\
\ch{SiH}  & Exomol & \citet{YurchenkoEtal2018mnrasExomolSiH}  \\
\ch{SiH4} & Exomol & \citet{OwensEtal2017mnrasExomolSiH4}  \\
\ch{SiO}  & Exomol & \citet{YurchenkoEtal2022mnrasSIOUVENIRexomolSiO}  \\
\ch{SiS}  & Exomol & \citet{UpadhyayEtal2018mnrasExomolSiS}  \\
\ch{SO}   & Exomol & \citet{BradyEtal2024mnrasExomolSO}  \\
\ch{SO2}  & Exomol & \citet{UnderwoodEtal2016mnrasSO2exoamesExomol}  \\
\ch{TiO}  & Exomol & \citet{McKemmishEtal2019mnrasTOTOexomolTiO}  \\
\ch{VO}   & Exomol & \citet{BowesmanEtal2024ExomolVOhyvo}  \\
\hline
\end{tabular*}
\end{table}
}

\subsection{Fitting data to models}

\subsubsection{Depth offsets}

The analysis of multi-epoch exoplanet time-series with {\JWST} has
revealed transit-depth offsets between observations at overlapping
wavelengths on the order of 10--100~ppm.  These offsets have been
attributed to astrophysical effects (e.g., stellar variability,
unocculted stellar spots), data-reduction choices (e.g., adopted
system parameters, limb-darkening treatment), and instrumental
systematics \citep{MadhusudhanEtal2023apjK2-18bJWSTcarbonMolecules,
  CarterEtal2024natasDataSynthesisWASP39b,
  FuEtal2024natasHD189733bH2S,
  WelbanksEtal2024natWASP107bJWSTtransits,
  BaratEtal2025ajV1298TaubTransitNIRSpec,
  LouieEtal2025ajWASP17bTransitSOSS,
  MayoEtal2025ajWASP166bTransitJWST}.  Furthermore, depth offsets have
been seen between detectors during a same observation
\citep[e.g.,][]{FournierTondreauEtal2025mnrasWASP52bTransitJWST,
  HolmbergMadhusudhanEtal2024aaTOI270dJWSThycean}.  These offsets are
unlikely to be of astrophysical origin.  A variety of potential
explanations have been suggested, such as different background levels
between detectors
\citep{GressierEtal2024apjL98-59dTransitJWSTsulphur}, systematics due
to the small number of groups per integration
\citep{AdamsRedaiEtal2025apjGJ357b}, performance or use differences
between detectors \citep{AhrerEtal2025mnrasWASP94AbTransitNIRSpec},
treatment of the bias correction
\citep{MoranEtal2023apjGJ486bJWSTwaterOrStar}, differences in the 1/f
noise subtraction \citep{LothringerEtal2025ajWASP178bNIRSpecTransit},
or light-curve detrending analysis
\citep{BelloArufeEtal2025apjVolcanismL98-59b,
  SikoraEtal2025ajHD80606bSeasonsG395H}.

To account for possible offsets between observations or detectors, we
have implemented parametric models that shift the depths of selected
data points.  These are wavelength-independent offsets that can be applied
to individual or groups of data points, i.e., to specific observations
or instruments.

\subsubsection{Noise scaling}

{\JWST} spectra often show noise levels above that
expected from pure instrumental noise or residual correlated noise
\citep[e.g.,][]{EspinozaEtal2023paspNIRSpecCommissioning,
  ScarsdaleEtal2024ajL98-59cNIRSpecTransit,
  AldersonEtal2024ajTOI836bTransitG395H,
  WallackEtal2024ajTOI836cTransitG395H,
  AlamEtal2025ajL168-9bTransitsNIRSpecMIRI,
  MayoEtal2025ajWASP166bTransitJWST}.  This suggest that the structure
and magnitude of this noise source is not well understood yet
\citep{BellEtal2024natasWASP43bMIRIphase}.

To account for possible underestimated uncertainties, we implemented
parametric models that scale the original depth uncertainties
($\sigma_i$), either as multiplicative factors
\citep{BellEtal2024natasWASP43bMIRIphase} or added in quadrature
\citep{LineEtal2015apjBrownDwarfRetrievalsI,
  WelbanksMadhusudhan2021apjAurora}, respectively:
\begin{align}
\sigma'_i  & =  S \sigma_i, \\
\indent \sigma'_i & =  \sqrt{\sigma_i^2 + S^2},
\end{align}
where $\log S$ is the model parameter.  As before, the error-scaling
models are wavelength independent, and can apply to individual or
groups of data points.

\subsection{Transit light source effect}

The transit light source (TLS) effect is a wavelength-dependent
contamination into exoplanet transmission spectra due to spots or
faculae heterogeneities on the surface of the host star
\citep{McCulloughEtal2014apjHD189733bHST,
  RackhamEtal2018apjStellarHeterogeneityI,
  ThompsonEtal2024apjStellarActivityCorrection}.  Stellar activity has
become a recurrent source of noise and bias in atmospheric
characterization studies from {\JWST} observations, in particular for
M and K stellar-type hosts or young stars
\citep[e.g.,][]{MoranEtal2023apjGJ486bJWSTwaterOrStar,
  AhrerEtal2025apjGJ3090bHeliumSOSS,
  MurphyEtal2026ajV1298cTransitJWST}.
In {\pyratbay} we modeled the TLS correction for unocculted spots following
\citet{RackhamEtal2018apjStellarHeterogeneityI}, where the observed
transit depth is expressed in terms of the true transit depth as
$d_{\rm obs}(\lambda) = \epsilon(\lambda) d(\lambda)$, with the TLS
correction given by:
\begin{equation}
\epsilon(\lambda) = \frac{1}{
  1 - f_{\rm spot}\left(1-{F_{\rm spot}} / {F_s}\right)
    - f_{\rm fac}\left(1-{F_{\rm fac}} / {F_s}\right)
},
\end{equation}
where $F_s$, $F_{\rm spot}$, and $F_{\rm fac}$ are the flux spectra
from the uncontaminated host star, spots, and faculae, respectively.
$f_{\rm spot}$ and $f_{\rm fac}$ are the spot and faculae coverage
fraction of the stellar surface (projected area) as seen from Earth at
the moment of the observation.  These fluxes are modeled using a
library of SED models covering the expected range of stellar, spot,
and faculae temperatures.  {\pyratbay} uses the PHOENIX New-Era SED
models \citep{HauschildtEtal2025aaPHOENIXnewEra}, which typically span
temperatures from $T \approx 2300$~K to $12000$~K, although these
boundaries may vary depending on the stellar metallicity and surface
gravity.  The model parameters are thus the spot (and optionally
faculae) temperature and covering fraction.

\begin{figure}
\includegraphics[width=\linewidth,clip]{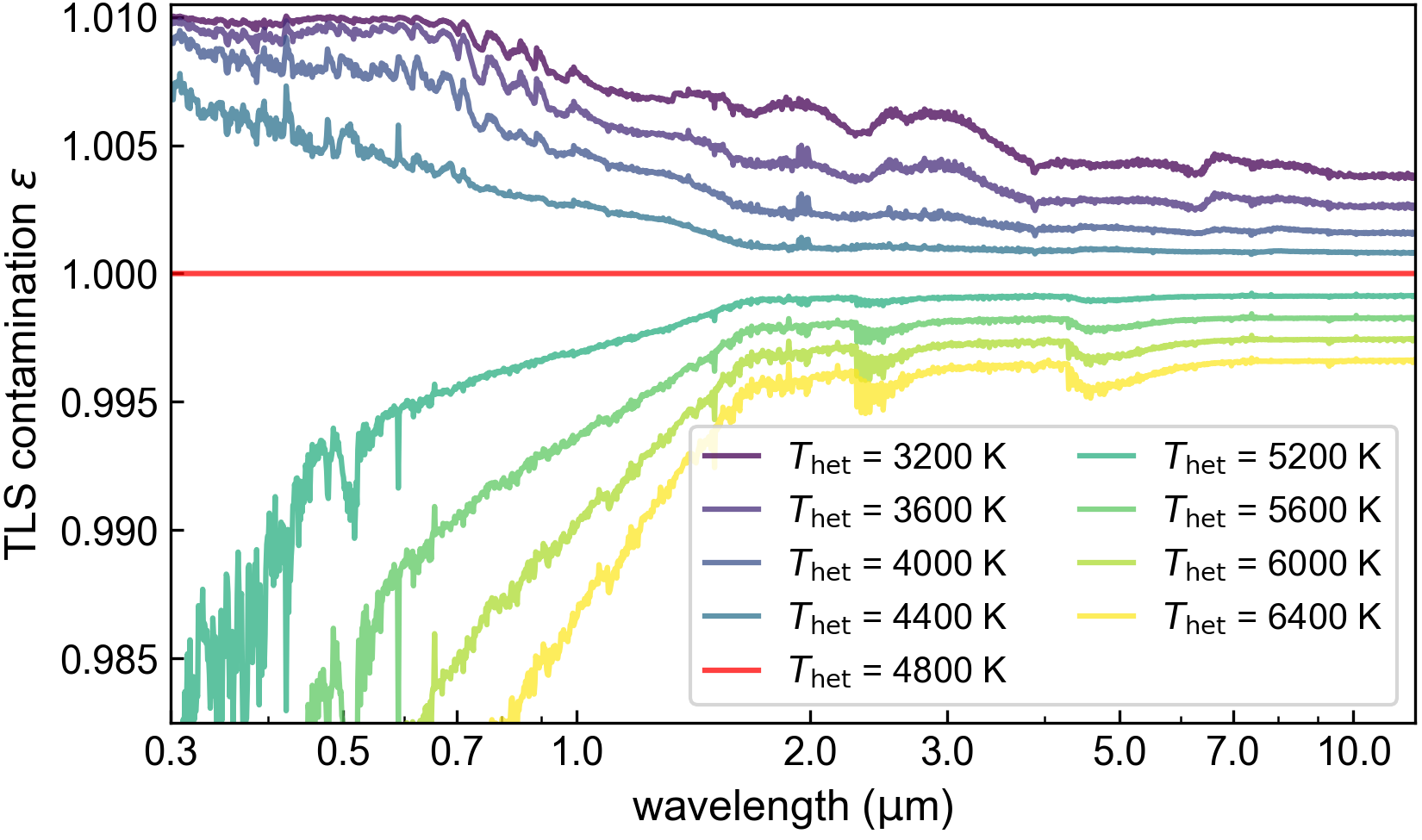}
\caption{Transit light source (TLS) effect for a [M/H]=0.5 K2V star with
  $T_{\rm eff}=4800$~K and $\log g=4.5$.  Each curve shows the
  contamination to the optical--NIR transmission spectrum produced by a spot
  ($T_{\rm het} < T_{\rm eff}$) or faculae ($T_{\rm het} > T_{\rm
    eff}$) with a projected covering fraction of 1\% at a temperature
  $T_{\rm het}$ (see legend).}
\label{fig:tls_effect}
\end{figure}

Figure \ref{fig:tls_effect} shows an example of the TLS effect on a
[M/H]=0.5 K2V star ($T_{\rm eff}=4800$~K and $\log g=4.5$) produced by
spot/faculae heterogeneities with a covering fraction of $f=0.01$ over
a range of temperatures.  The TLS contamination is the largest at
short wavelengths ($\lambda\lesssim2$), exhibiting a strong spectral
component.  At longer wavelengths the amplitude of the TLS
contamination decreases, although significant chromatic variability
can remain up to $\sim$10~{\micron}.

Given that spots and faculae come into and out of view as a host star
rotates, stellar heterogeneities can induce flux variability in
timescales as short as a few days.  To model this time dependency,
{\pyratbay} can selectively apply individual TLS corrections to
specific groups of data points, enabling simultaneous atmospheric
fitting and retrieval from multi-epoch observations.

\begin{figure*}
\includegraphics[width=\linewidth]{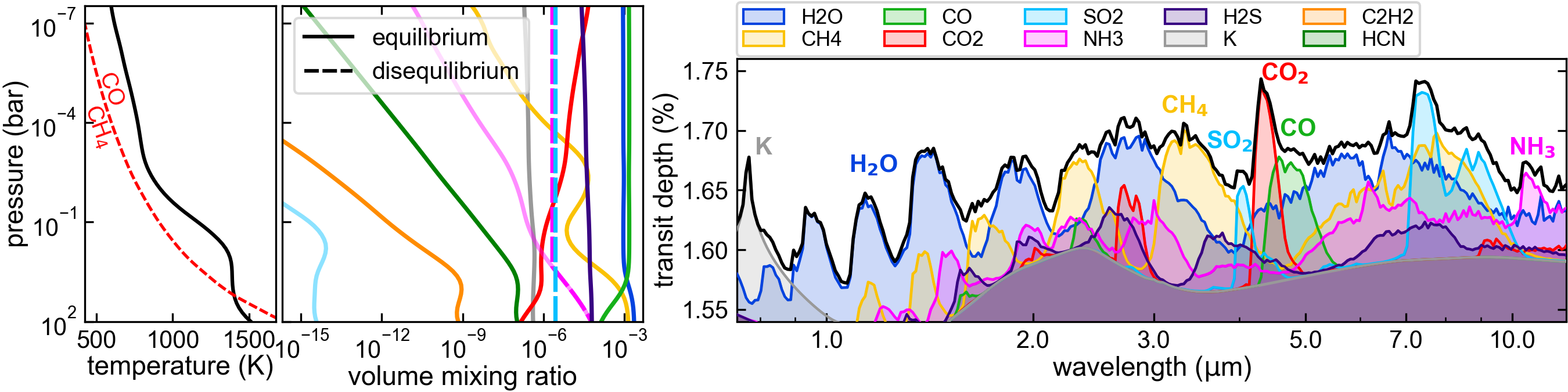}
\caption{{\bf Left panels:} radiative-thermochemical-equilibrium
  simulation of {\sixtynineb} assuming a 3$\times$ solar metallicity
  and a solar C/O ratio of 0.59.  The dashed red curve in the left
  panel shows the boundary where thermochemical equilibrium predicts
  an equal amount of CO and \ch{CH4}. Volume mixing ratios are color
  coded according to the legends above the right panel. The dashed VMR
  curves show \ch{SO2} and \ch{NH3} abundances taken out of
  equilibrium. {\bf Right panel:} transmission spectrum of the
  {\sixtynineb} model atmospheres (black curve).  The colored curves
  with shaded areas show the contribution from specific atmospheric
  species to the transmission spectrum (see legend and labels).  }
\label{fig:WASP69b_atmosphere_spectra}
\end{figure*}

\section{Application: Characterizing exoplanet atmospheres with {\JWST}}
\label{sec:applications}

The unprecedented signal-to-noise ratio and spectral coverage of
{\JWST} exoplanet observations is enabling access to the
three-dimensional distribution of atmospheric properties. Thanks to
the detection of limb asymmetries between morning and evening
terminators in several giant planets
\citep[e.g.,][]{EspinozaEtal2024natWASP39bInhomogeneousTerminators,
  MurphyEtal2024natasLimbAsymmetryWASP107b}, much focus has been
placed at the study of longitudinal variability.
In this section, we complement this picture, investigating {\JWST}'s
potential to constrain vertical variations in atmospheric abundances,
their dependence on spectral coverage, and the biases that may arise
if these variations are not properly accounted for.

To demonstrate the capability of the new tools developed for
{\pyratbay}, we performed atmospheric retrieval simulations with {\JWST}
on synthetic transmission spectra of the warm-Jupiter {\sixtynineb}
\citep{AndersonEtal2014mnrasWASP84b_69b_70Ab}.  We selected
this planet because its large radius and relative low mass \citep[1.06
  {\rjup}, 0.25 {\mjup},][]{BonomoEtal2017aaRVmasses}, combined with
its bright host star (K2V, 7.46 K$_s$ mag) make {\sixtynineb} a
favorable target for atmospheric characterization
\citep{KemptonEtal2018paspTransitSpectroscopicMetric}.  {\sixtynineb}
is also being observed by {\JWST} with NIRISS/SOSS, NIRSpec/G395H, and
MIRI/LRS (PIDs
\href{https://www.stsci.edu/jwst/science-execution/program-information?id=3712}{3712}
and
\href{https://www.stsci.edu/jwst/science-execution/program-information?id=5924}{5924}),
yielding a combined 0.8--12~{\microns} transmission spectrum that
probes multiple absorption features by \ch{H2O}, \ch{CH4}, CO,
\ch{CO2}, \ch{NH3}, or \ch{SO2}.
Thus, synthetic spectra for such a target provide a control dataset to
compare input and retrieved atmospheric properties, and test how the
posterior constraints depend on the spectral coverage.

\subsection{Atmospheric forward modeling of {\sixtynineb}}

We based our simulations on the observational constraints of
{\sixtynineb} from the literature.  Ground-based observations with the
high-resolution instrument GIANO-B (0.9--2.45 {\micron}, at
$R=50\,000$) detected the signature of \ch{H2O}, CO, \ch{CH4},
\ch{NH3}, and \ch{C2H2} \citep{GuilluyEtal2022aaGianoWASP69b}.
Optical and near-infrared observations have consistently detected
\ch{H2O} at solar-to-supersolar abundances, and a transit-depth slope
characteristic of non-gray hazes
\citep{FisherHeng2018mnrasWFC3SampleRetrieval,
  TsiarasEtal2018ajPopulationStudy, MurgasEtal2020aaWASP69bGTC,
  KhalafinejadEtal2021aaWASP69bHighResTransit}.

We generated temperature and VMR profiles under
radiative-thermochemical equilibrium, assuming efficient day-to-night
energy redistribution, zero bond albedo, 3$\times$ solar metallicity,
and a solar carbon-to-oxygen ratio of C/O = 0.59.  The computed
temperature profile is non-inverted, spanning 600--1400~K between
$10^{-5}$ and 10~bar (Fig.\ \ref{fig:WASP69b_atmosphere_spectra}, left
panel).  At these temperatures, CO is the primary carbon carrier,
leaving a lower abundance of \ch{CH4}.  With a solar C/O ratio, the
excess oxygen not bound to CO forms primarily \ch{H2O}
(Fig.\ \ref{fig:WASP69b_atmosphere_spectra}, middle panel).  Motivated
by the detection of \ch{NH3}, suggesting disequilibrium chemistry \citep{GuilluyEtal2022aaGianoWASP69b}, we
manually enhanced the abundance of \ch{NH3} and \ch{SO2} in our models
to mimic the effects of vertical quenching and photochemistry
\citep[e.g.,][]{Moses2014rsptaChemicalKinetics,
  TsaiEtal2023natWASP39bPhotochemistry}.

We computed the {\sixtynineb} transmission spectrum
including line-sampled cross-sections for CO, \ch{CO2}, \ch{H2O},
\ch{CH4}, \ch{C2H2}, \ch{SO2}, \ch{H2S}, HCN, \ch{NH3} (Table
\ref{table:cross_sections}); alkali opacity for K
\citep{BurrowsEtal2000apjBDspectra}; collision-induced absorption for
\ch{H2}--\ch{H2} and \ch{H2}--He \citep{BorysowEtal1988apjH2HeRT,
  BorysowEtal1989apjH2HeRVRT, BorysowEtal2001jqsrtH2H2highT,
  BorysowFrommhold1989apjH2HeOvertones, Borysow2002jqsrtH2H2lowT}; and
Rayleigh-scattering for \ch{H2} and He \citep{Kurucz1970saorsAtlas}.
The transmission spectrum (Fig. \ref{fig:WASP69b_atmosphere_spectra},
right panel) illustrates the broad range of atmospheric species
accessible to near-infrared transmission observations.  In particular,
species like \ch{H2O} and \ch{CH4} present multiple absorption
features in this range, potentially arising from different pressure
levels, which is key to enable constraints of their VMR vertical
profiles.

\begin{figure*}
\includegraphics[width=\linewidth]{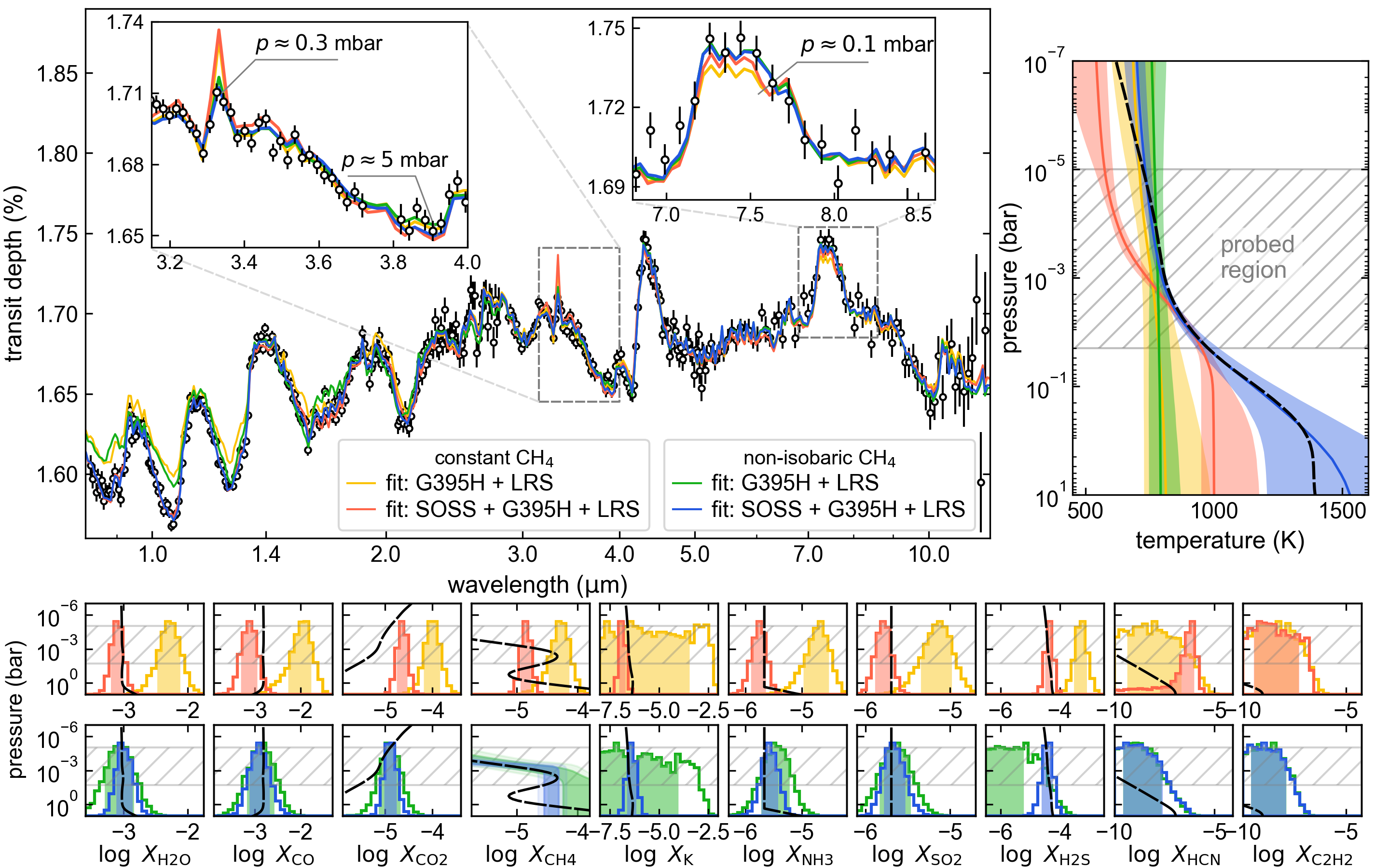}
\caption{Retrieval of simulated {\JWST} transmission spectra of
  {\sixtynineb}. {\bf Top left:} the black markers with error bars
  show the simulated {\JWST} spectra with NIRISS, NIRSpec, and MIRI.
  The solid curves show the retrieved spectra, color coded by fitting
  dataset and VMR parameterization (see legend). The insets zoom in
  on prominent \ch{CH4} absorption bands, arising from different
  pressure levels. {\bf Top right:} posterior pressure--temperature
  profiles.  The shaded areas denote the retrieved 1$\sigma$ credible
  intervals (same color coding as left panel), whereas the black
  dashed curve shows the input true temperature profile.  The hatched
  area marks the pressure ranges probed by the simulated transmission
  spectra. {\bf Bottom:} marginal posterior distributions for the
  atmospheric abundances.  The shaded areas denote the 1$\sigma$
  credible intervals for each retrieval run (same color coding as
  above).  The black dashed curves show the true values as function of
  pressure.  The panels for \ch{CH4} shows the abundance profile as a
  function of pressure.  The hatched region denotes the range of
  pressures probed by the simulated observations.}
\label{fig:WASP69b_retrieval}
\end{figure*}

\subsection{Atmospheric retrievals of {\sixtynineb}}

To simulate {\JWST} transmission observations of our model spectra we
employed the \textsc{Gen TSO} simulator for time-series observations
\citep{Cubillos2024paspGenTSO}.  We simulated a single transit
observation with each instrument of the GO programs 3712 and 5924.  We
computed the expected noise following the instrumental configuration
of these programs and assuming a K2V PHOENIX SED for WASP-69 ($T_{\rm
  eff}=4750$~K, $\log g=4.5$). We then increased the noise by a factor
of 1.5 to account for the excess noise typically seen in observed data.  At a
resolution of $R=100$, the mean precision of the transit depths are
30~ppm for NIRISS/SOSS (0.8--2.8 {\microns}), 33~ppm for NIRSpec/G395H
(2.9--5.2 {\microns}), and 84~ppm for MIRI/LRS (5--12 {\microns}).
Figure \ref{fig:WASP69b_retrieval} shows the simulated spectra.

For the retrievals, we adopted the commonly-used ``free-chemistry''
parameterization to explicitly assess its ability to constrain the
composition of more complex atmospheric models.  We focused on
\ch{CH4} because it is the most abundant trace gas showing a
pressure-dependent VMR.  Also, \ch{CH4} presents multiple spectral
features at 2.2, 3.3, and 8.0 {\microns}.  Thus, for all other species
we assumed constant-with-altitude VMR profiles, whereas for \ch{CH4}
we tested both a constant and a non-isobaric VMR profile.  This
comparison allows us to determine compositional biases, and their
statistical significance.
We also explored the impact of spectral coverage on the abundance
constraints by comparing retrievals using all three instruments
(0.8--12~{\microns}) to retrieval using only NIRSpec and MIRI
(2.8--12~{\microns}).

The retrievals have in total 18 (for the constant-VMR for \ch{CH4}
case) and 20 (non-isobaric \ch{CH4} VMR case) free parameters.  We fit
the VMR of \ch{H2O}, \ch{CO}, \ch{CO2}, \ch{CH4}, K, \ch{NH3},
\ch{SO2}, \ch{H2S}, \ch{HCN}, and \ch{C2H2}.  Simultaneously, we fit
the temperature profile with a
\citet{MadhusudhanSeager2009apjRetrieval} model, the reference
pressure of the planet radius, and the cloud-top pressure $\pcloud$ of
a gray cloud-deck model.  We carried out the posterior sampling with
{\multinest}, using 1000 live samples.

Table \ref{table:WASP69b_retrieval} lists the models free parameters,
their priors, and the retrieved values.  Figure
\ref{fig:WASP69b_retrieval} shows the retrieved spectra and posteriors
for each model parameterization and fitting dataset.  Overall, these
simulated transmission spectra probe pressures from
$10^{-2}-10^{-5}$~bar, range in which the \ch{CH4} abundance decreases
steeply with altitude.

The NIRSpec+MIRI retrieval assuming all constant VMRs fails to
reproduce all spectral features (Fig.\ \ref{fig:WASP69b_retrieval},
yellow model and posteriors).  The core of the discrepancy arises from
the constant-VMR assumption for \ch{CH4}, as highlighted in
the transmission spectrum insets.
These insets show two spectral regions with prominent \ch{CH4}
features, but that probe distinct pressure levels.  While a
constant \ch{CH4} profile can fit the spectrum at $\sim$3.9~{\micron}
(probing 5~mbar, where \ch{CH4} is abundant), is not able to
simultaneously fit the 3.3~{\micron} feature (probing 0.03~mbar, where
\ch{CH4} is less abundant).  Furthermore, given the degenerate nature
of atmospheric abundance retrievals, correlations between abundances
propagate into biases for other species, resulting in overestimated
VMR posteriors.
In particular, the VMR constraints for \ch{H2O} and \ch{CO}, the two
most abundant trace gasses, are $\sim$3.5$\sigma$ above the true
values, leading to a biased metallicity inference of $\sim$18$\times$
solar, rather than the expected 3$\times$ solar value.
We found that the cloud top-pressure parameter is constrained to the
layers below the observed photosphere, showing no significant impact
on the retrievals nor degeneracies with other parameters. This is
consistent with the input cloud-free atmospheric model.

The addition of NIRISS/SOSS data provides spectral coverage of several
\ch{H2O} bands in the 0.8--2.8~{\micron} range, which improves the
overall accuracy of the constant-VMR constraints
(Fig.\ \ref{fig:WASP69b_retrieval}, red model).  The retrieved
\ch{CH4} VMR falls more in line with the average true-VMR (between the
probed pressures), but again the model is unable to fit all spectral
features (see insets).  Consequently, biases emerge in the form of
lower temperatures than expected, a spurious HCN posterior peak at
$\log X_{\rm HCN} =-6.4$, and a slight overestimation of \ch{CO2}.

We then performed atmospheric retrievals using the non-isobaric VMR
model for \ch{CH4}.  These runs produced substantially better fits to
the simulated spectra (Fig.\ \ref{fig:WASP69b_retrieval}, green and
blue).  The resulting posterior distributions accurately recovered the
true VMR values within 1$\sigma$ credible intervals for detected
species, and yielded upper limits for non-detected species.  The
wavelength coverage played a smaller factor compared to the
constant-VMR retrievals.  In this case, adding NIRISS data extends the
range of probed pressures, improving the overall constraints.  That
is, narrower posterior distributions, a more accurate fit of the
non-inverted thermal profile, and a strengthened \ch{CH4} constraint
at high pressures, due to the inclusion of the \ch{CH4} band at
1.7~{\micron}.

Finally, we note that although the \ch{CO2} abundance also varies with
altitude, the retrievals (assuming constant \ch{CO2} VMR profiles) do
not appear to be significantly biased.  Two factors likely explain
this.  First, in this spectrum \ch{CO2} exhibits only two absorption
bands at 4.4 and 2.8~{\micron}, with the latter located at the
wavelengths with the lowest S/N of the simulated data.
Second, due to the observing geometry in transit observations, rays
that probe deeper layers of an atmosphere also traverse low-pressure
layers.  Therefore, for an abundance that increases with altitude,
like \ch{CO2}, the stronger absorption at low pressures dominates the
shape of spectral features, and thus is less likely to introduce
altitude-dependent biases.  Consequently, as seen in these
simulations, the retrievals fit a VMR close to that of the
lowest-pressure layers that are probed.

These simulations show that {\JWST} observations of Jupiter-sized
exoplanets have the potential to characterize the vertical structure
of their atmospheric composition.  These are challenging measurements
that require favorable targets with extended atmospheres and high S/N
observations.  Although our simulations assumed somewhat idealized,
cloud-free conditions, metrics such as $\chi^2$ statistics or Bayes
factor ($B$) help determine when a more complex model is justified in
a robust manner \citep[e.g.,][]{Raftery1995BIC, Trotta2007mnrasBayesianModelSelection, ThorngrenEtal2026apjsBayesModelComparison}.  For example, the
Bayes factor between competing models can be estimated from the ratio
of their posterior evidences, an output from {\multinest}.  In our
{\sixtynineb} retrieval simulations, the reduced $\chi^2$
  improves from 1.56 to 1.09 for the NIRSpec+MIRI fit when switching
  from the isobaric to non-isobaric VMR assumption. Likewise, for the
  NIRISS+NIRSpec+MIRI fit the reduced $\chi^2$ improves from 1.61 to
  1.14. The Bayesian evidence statistics also strongly favores the
non-isobaric over the constant-VMR model, with $\ln B=35.7$
for the NIRSpec+MIRI fit and $\ln B=89.1$ for the NIRISS+NIRSpec+MIRI
fit.
Lastly, we acknowledge the practical limitations of using more complex
models, as the larger dimensionality of the parameter space steeply
increases the number of model evaluations for the posterior sampling.
The use self-consistent VMR profiles, e.g., from kinetic or global
  circulation models, can help to determine when to consider
vertically varying VMR profiles for a given species.

\section{Conclusions}
\label{sec:discussion}

We have presented a major upgrade to the {\pyratbay} atmospheric
modeling package.  This includes a new self-standing
radiative-equilibrium code {\chemcat}, specifically designed for
exoplanet atmospheric modeling.  {\chemcat} enabled the implementation
of radiative-equilibrium and equilibrium-chemistry retrieval studies
in {\pyratbay}.  We have successfully validated these new modules by
comparing atmospheric models to existing chemistry and
radiative-transfer codes from the literature (\ggchem, \fastchem,
\helios).  We have also incorporated into {\pyratbay} state-of-the-art
retrieval schemes, including {\multinest} nested sampling,
isotopic-ratio fitting, equilibrium-chemistry retrievals, and hybrid
equilibrium-chemistry plus free-VMR retrievals.  We have also
implemented the transit-light-source effect to correct for stellar
contamination by unocculted spots and faculae in transmission spectra.
We further implemented fitting and retrieval parameters for data
manipulation, that is error-scaling and transit-depth offsets for
multi-epoch observations.  Lastly we developed a new abundance
free-chemistry parameterization allowing vertical variation in VMR
profiles.  We show with simulated transit observations that {\JWST} is
able to detect vertical variations.  Additionally, a separate work by
Blecic et al. (in prep.) will present the implementation of advanced
mie scattering clouds.  The need for such tools is evidenced by the
wide variety of research approaches seen in the literature, such as
atmospheric grid fitting for high-resolution observations
\citep{GiacobbeEtal2021natSixMoleculesCNO,
  CarleoEtal2022ajGianoWASP80b, GuilluyEtal2022aaGianoWASP69b,
  BasilicataEtal2024aaGianoHATP11b}, atmospheric retrievals of
ultra-hot Jupiters \citep[][]{DemangeonEtal2024aaWASP76bCHEOPS,
  SinghEtal2024aaCheopsKELT20b, AkinsanmiEtal2024aaWASP12bCHEOPS,
  ChallenerEtal2025natasWASP18bSOSSmapping,
  DelineEtal2025aaWASP18bCHEOPS, GaraiEtal20025aaKELT7bCHEOPS}, or
atmospheric simulations to study the characterization potential of
{\JWST} \citep[][]{BangeraEtal2025apjKinematicsWASP69b,
  BonfantiEtal2025aaTOI396systemCHEOPS, DamassoEtal2024aaTOI837b,
  SozzettiEtal2024mnrasK2370bTESSandCHEOPS}.

{\pyratbay} is a Python package available for installation from
PyPI\footnote{\href{https://pypi.org/project/pyratbay}
{https://pypi.org/project/pyratbay}} (\textsc{pip install pyratbay}).
The source code is available at
\href{https://github.com/pcubillos/pyratbay}
     {github.com/pcubillos/pyratbay}.
To facilitate the use of {\pyratbay}, we have provided via
Zenodo\footnote{\href{https://zenodo.org/records/16965390}
{https://zenodo.org/records/16965390}} pre-computed line-sampled cross
sections for 38 species of interest for exoplanet atmospheric
modeling.  These cross sections cover wavelength from 0.15 to
33~{\micron}, pressures from $10^3$ to $10^{-9}$~bar, and temperatures
from 200 to 5000~K, at a spectral resolution of $R=25,000$.  We have
also expanded the documentation, providing practical tutorials of
specific modules and full end-to-end scientific
applications\footnote{\href{https://pyratbay.readthedocs.io/en/latest/cookbooks/recipes.html}
{https://pyratbay.readthedocs.io/en/latest/cookbooks/recipes.html}}.

\section*{Acknowledgements}

We thank the anonymous referee for their time and valuable comments.
We thank Luis Welbanks and Jorge Naranjo for their help implementing nested sampling.
We thank contributors to the Python Programming Language and the free
and open-source community (see Software Section below).  This project
was funded by the Austrian Science Fund (FWF) Erwin Schroedinger
Fellowship, program J4595-N.
D.\ S.\ acknowledges support from the Severo Ochoa grant CEX2021-001131-S
funded by MCIN/AEI/10.13039/501100011033 and from the project
PID2021-126365NB-C21(MCI/AEI/FEDER, UE).
This research has made use of NASA's Astrophysics Data System
Bibliographic Services.

The following software and packages were used in this work:
{\pyratbay} \citep{CubillosBlecic2021mnrasPyratBay},
{\mcc} \citep{CubillosEtal2017apjRednoise},
{\multinest} \citep{FerozEtal2009mnrasMultiNest},
{\chemcat}\footnote{
\href{https://chemcat.readthedocs.io}
     {https://chemcat.readthedocs.io}},
\textsc{TEA} \citep{BlecicEtal2016apsjTEA},
\textsc{GGchem} \citep{WoitkeEtal2018aaGGchem},
\textsc{FastChem} \citep{StockEtal2018mnrasFastChem},
\textsc{HELIOS} \citep{MalikEtal2017ajHELIOS},
{\repack} \citep{Cubillos2017apjRepack},
\textsc{Numpy} \citep{HarrisEtal2020natNumpy},
\textsc{SciPy} \citep{VirtanenEtal2020natmeScipy},
\textsc{Matplotlib} \citep{Hunter2007ieeeMatplotlib},
\textsc{IPython} \citep{PerezGranger2007cseIPython},
and
\textsc{bibmanager}\footnote{
\href{https://bibmanager.readthedocs.io}
     {https://bibmanager.readthedocs.io}}
\citep{Cubillos2020zndoBibmanager}.

\section*{Data Availability}

Data and analyses presented in this article are available at
this Zenodo research compendium
\href{https://doi.org/10.5281/zenodo.21689779}
     {https://doi.org/10.5281/zenodo.21689779}.


\bibliographystyle{mnras}
\bibliography{radeq}


\appendix

\section{{\sixtynineb} retrieval parameters}
Table \ref{table:WASP69b_retrieval} lists all parameters, priors, and
posteriors for the atmospheric retrievals of the simulated WASP-69b
spectra.

{\renewcommand{\arraystretch}{1.2}
\begin{table*}
\centering
\caption{Atmospheric retrievals of simulated WASP-69\,b observations}
\label{table:WASP69b_retrieval}
\strut\hfill
\begin{tabular} {lccccccc}
\hline
\hline
                               & Uniform prior    &  \multicolumn{2}{c}{Constant \ch{CH4} retrievals}             &  \multicolumn{2}{c}{Non-isobaric \ch{CH4} retrievals}   \\
Parameter                      & ranges           &  NIRSpec+MIRI                 & NIRISS+NIRSpec+MIRI           &  NIRSpec+MIRI              & NIRISS+NIRSpec+MIRI \\
\hline
$\log\ p_1$ (bar)              & $(-9.0, 2.0)$    &  $-4.4^{+3.0}_{-3.3}$         &  $-3.57^{+2.15}_{-1.00}$      &  $-6.1^{+3.0}_{-2.0}$      &  $-2.51^{+0.54}_{-0.73}$ \\
$\log\ p_2$ (bar)              & $(-9.0, 2.0)$    &  $-4.5^{+3.9}_{-3.0}$         &  $-6.2^{+2.2}_{-1.8}$         &  $-3.7^{+3.4}_{-3.2}$      &  $-5.9^{+2.1}_{-2.1}$    \\
$\log\ p_3$ (bar)              & $(-9.0, 2.0)$    &  $-0.7^{+1.9}_{-3.0}$         &  $-1.61^{+2.16}_{-0.49}$      &  $-1.8^{+2.6}_{-3.4}$      &  $0.4^{+1.1}_{-1.1}$     \\
$a_1$                          & $(0.02, 2.0)$    &  $1.23^{+0.49}_{-0.43}$       &  $1.44^{+0.38}_{-0.74}$       &  $1.31^{+0.46}_{-0.54}$    &  $1.39^{+0.39}_{-0.33}$  \\
$a_2$                          & $(0.02, 2.0)$    &  $1.21^{+0.52}_{-0.55}$       &  $0.49^{+0.39}_{-0.16}$       &  $1.36^{+0.43}_{-0.50}$    &  $0.42^{+0.14}_{-0.12}$  \\
$T_0$ (K)                      & $(300, 1500)$    &  $674.6^{+104.8}_{-94.4}$     &  $531.8^{+43.0}_{-155.2}$     &  $748.4^{+70.3}_{-62.1}$   &  $688.5^{+36.5}_{-56.3}$ \\
$\log p_{{\rm ref}}$ (bar)     & $(-8.0, 0.0)$     &  $-5.99^{+0.23}_{-0.21}$      &  $-5.233^{+0.093}_{-0.092}$   &  $-5.02^{+0.21}_{-0.22}$   &  $-5.04^{+0.11}_{-0.13}$ \\
$\log\ X_{\rm H2O}$            & $(-10.0, -0.5)$  &  $-2.29^{+0.18}_{-0.19}$      &  $-3.099^{+0.094}_{-0.100}$   &  $-3.06^{+0.22}_{-0.23}$   &  $-2.99^{+0.14}_{-0.12}$ \\
$\log\ X_{\rm CO}$             & $(-10.0, -0.5)$  &  $-1.98^{+0.23}_{-0.29}$      &  $-3.12^{+0.17}_{-0.18}$      &  $-2.86^{+0.29}_{-0.31}$   &  $-2.91^{+0.23}_{-0.21}$ \\
$\log\ X_{\rm CO2}$            & $(-10.0, -0.5)$  &  $-4.01^{+0.17}_{-0.18}$      &  $-4.64^{+0.11}_{-0.12}$      &  $-4.90^{+0.23}_{-0.23}$   &  $-4.89^{+0.14}_{-0.13}$ \\
$\log\ X_{\rm CH4}$            & $(-10.0, -0.5)$  &  $-4.29^{+0.16}_{-0.19}$      &  $-4.847^{+0.084}_{-0.089}$   &  $\cdots$                  &  $\cdots$                \\
$m_{\rm CH4}$                  & $(0.0, 5.0)$     &  $\cdots$                     &  $\cdots$                     &  $1.22^{+0.92}_{-0.27}$    &  $2.6^{+1.5}_{-1.1}$     \\
$\log\ X_{\rm CH4}^{0}$        & $(-10.0, 1.0)$   &  $\cdots$                     &  $\cdots$                     &  $-4.45^{+0.63}_{-0.41}$   &  $-3.93^{+0.84}_{-0.58}$ \\
$\log\ X_{\rm CH4}^{\rm max}$  & $(-6.0, -0.5)$   &  $\cdots$                     &  $\cdots$                     &  $-3.94^{+1.96}_{-0.29}$   &  $-4.43^{+0.13}_{-0.11}$ \\
$\log\ X_{\rm K}$              & $(-10.0, -0.5)$  &  $-6.1^{+2.7}_{-2.6}$         &  $-6.85^{+0.16}_{-0.17}$      &  $-6.5^{+2.4}_{-2.2}$      &  $-6.39^{+0.24}_{-0.23}$ \\
$\log\ X_{\rm NH3}$            & $(-10.0, -0.5)$  &  $-4.72^{+0.20}_{-0.26}$      &  $-5.75^{+0.12}_{-0.14}$      &  $-5.46^{+0.25}_{-0.26}$   &  $-5.57^{+0.15}_{-0.15}$ \\
$\log\ X_{\rm SO2}$            & $(-10.0, -0.5)$  &  $-5.15^{+0.17}_{-0.21}$      &  $-5.789^{+0.093}_{-0.096}$   &  $-5.64^{+0.18}_{-0.19}$   &  $-5.66^{+0.14}_{-0.12}$ \\
$\log\ X_{\rm H2S}$            & $(-10.0, -0.5)$  &  $-3.30^{+0.21}_{-0.23}$      &  $-4.36^{+0.13}_{-0.13}$      &  $-7.1^{+1.8}_{-1.9}$      &  $-4.42^{+0.19}_{-0.19}$ \\
$\log\ X_{\rm HCN}$            & $(-10.0, -0.5)$  &  $-8.0^{+1.4}_{-1.3}$         &  $-6.37^{+0.35}_{-1.33}$      &  $-8.66^{+1.03}_{-0.90}$   &  $-8.64^{+1.06}_{-0.93}$ \\
$\log\ X_{\rm C2H2}$           & $(-10.0, -0.5)$  &  $-8.4^{+1.1}_{-1.1}$         &  $-8.5^{+1.2}_{-1.0}$         &  $-8.78^{+0.89}_{-0.83}$   &  $-8.79^{+0.89}_{-0.81}$ \\
$\log\ p_{\rm cl}$ (bar)       & $(-7.0, 2.0)$  &  $-0.1^{+1.4}_{-1.4}$         &  $0.87^{+0.73}_{-0.73}$       &  $-0.0^{+1.3}_{-1.4}$      &  $0.58^{+0.95}_{-0.98}$  \\
\hline
\end{tabular}
\hfill\strut
\begin{tablenotes}
\item[1]  The reported retrieved values correspond to the
 marginal posterior distribution's median and boundaries of the 68\%
 central credible interval \citep{Andrae2010arxivErrorEstimation}.
\item[2] For the non-isobaric \ch{CH4} model, the pressure of $\log\ X_{\rm CH4}^{0}$ is set at $p_0 = 1$ mbar.
\end{tablenotes}
\end{table*}
}


\bsp	
\label{lastpage}
\end{document}